\documentclass[aps,prb,amsmath, amssymb, english, twocolumn,reprint, floatfix, superscriptaddress]{revtex4-2}

\usepackage[mode=buildnew]{standalone}% requires -shell-escape
\usepackage[caption=false]{subfig}
\usepackage{bbm}
\usepackage{comment}
\usepackage{array}                                                                          
\usepackage{hhline}
\usepackage{tikz}
\usepackage{pgfplots}
\usepackage{microtype}
\usepackage{dsfont}
\usepackage{epstopdf}
\usepackage{epsfig} 
\usepackage[applemac]{inputenx}     
\usepackage{amsmath}
\usepackage{MnSymbol}
\usepackage{hyperref}

\pgfplotsset{compat=1.15}

\hypersetup{
    colorlinks,
    citecolor=blue,
    filecolor=blue,
    linkcolor=blue,
    urlcolor=blue
}

\newcommand{\bq}{{\mathbf{q}}}

\newcommand{\bk}{\mathbf{k}}

\newcommand{\bl}{{\boldsymbol\lambda}}

\usepackage{mathtools}

\newcommand{\SumCoreOmega}{\smashoperator[l]{\sum_{\omega\in\Omega^\nu_\text{c}}}}
\newcommand{\SumAsymOmega}{\smashoperator[lr]{\sum_{\omega\in\Omega^\nu_\text{s}}}}
\newcommand{\SumCoreNu}{\smashoperator[l]{\sum_{\nu\in\Omega^\omega_\text{c}}}}
\newcommand{\SumAsymNu}{\smashoperator[lr]{\sum_{\nu\in\Omega^\omega_\text{s}}}}
\newcommand{\SumCoreNup}{\smashoperator[l]{\sum_{\nu'\in\Omega^\omega_\text{c}}}}
\newcommand{\SumAsymNup}{\smashoperator[lr]{\sum_{\nu'\in\Omega^\omega_\text{s}}}}
\newcommand{\SumCoreNuNup}{\smashoperator[lr]{\sum_{\nu\nu'\in \Omega^\omega_\text{c}}}}
\newcommand{\SumAsymNuNup}{\smashoperator[lr]{\sum_{\nu\nu'\in \Omega^\omega_\text{s}}}}

\newcommand{\chiZeroShell}{\chi^{\mathrm{shell},\omega}_{0,\bq}}

\newcommand{\chiFullCore}{\chi^{\mathrm{core},\omega}_{r,\bq}}
\newcommand{\chiFullShell}{\chi^{\mathrm{shell},\omega}_{r,\bq}}

\newcommand{\lambdaFullCore}{\lambda^{\mathrm{core},\nu\omega}_{r,\bq}}
\newcommand{\lambdaFullShell}{\lambda^{\mathrm{shell},\nu\omega}_{r,\bq}}

\newcommand{\diff}{\mathrm{d}}
\newcommand{\magn}{\mathrm{m}}
\newcommand{\dens}{\mathrm{d}}
\newcommand{\lmagn}{{\lambda_\mathrm{m}}}
\newcommand{\ldens}{{\lambda_\mathrm{d}}}
\newcommand{\cmagn}{{\chi_\mathrm{m}}}

\newcommand{\smagnF}{{\Sigma^\nu_{\mathrm{m},\bk}}}
\newcommand{\sdensF}{{\Sigma^\nu_{\mathrm{d},\bk}}}
\newcommand{\cmagnF}{{\chi^{\omega}_{\mathrm{m},\bq}}}
\newcommand{\cdensF}{{\chi^{\omega}_{\mathrm{d},\bq}}}
\newcommand{\TN}{T_\mathrm{N}}
\newcommand{\lamm}{lD$\Gamma$A$_\text{m}\,$}
\newcommand{\lamdm}{lD$\Gamma$A$_\text{dm}\,$}

\def\equationautorefname~#1\null{Eq.~(#1)\null}

\allowdisplaybreaks

\begin{document}
\title{Consistency of potential energy in the dynamical vertex approximation}
\author{Julian Stobbe}
%\affiliation{Institute of Theoretical Physic, University of Hamburg, 20355 Hamburg, Germany}
\author{Georg Rohringer}
\affiliation{I. Institute of Theoretical Physics, University of Hamburg, 20355 Hamburg, Germany}

\date{\today} 

\pacs{}
\begin{abstract}

In the last decades, dynamical mean-field theory (DMFT) and its diagrammatic extensions have been successfully applied to describe local and nonlocal correlation effects in correlated electron systems.
Unfortunately, except for the exact solution, it is impossible to fulfill both the Pauli principle and conservation laws at the same time.
Consequently, fundamental observables such as the kinetic and potential energies are ambiguously defined.
In this work, we propose an approach to overcome the ambiguity in the calculation of the potential energy within the ladder dynamical vertex approximation (D$\Gamma$A) by introducing an effective mass renormalization parameter in both the charge and the spin susceptibility of the system.
We then apply our method to the half-filled single-band Hubbard model on a three-dimensional bipartite cubic lattice.
We find that:
(i) at weak-to-intermediate coupling, a reasonable modification of the transition temperature $\TN$ to the antiferromagnetically ordered state with respect to previous ladder D$\Gamma$A calculations without charge renormalization.
This is in good agreement with dual fermion and Monte Carlo results;
(ii) the renormalization of charge fluctuations in our new approach leads to a unique value for the potential energy which is substantially lower than corresponding ones from DMFT and non-self-consistent ladder D$\Gamma$A; and
(iii) the hierarchy of the kinetic energies between the DMFT and the ladder D$\Gamma$A in the weak coupling regime is restored by the consideration of charge renormalization.

\end{abstract}
\maketitle

\section{Introduction}
\label{sec:Intro}
The description and understanding of interacting many-particle systems represents one of the fundamental challenges in contemporary physics.
It arises in various research areas which include nuclear and atomic physics~\cite{Hen2017}, solid state theory~\cite{Held2007}, or soft matter systems~\cite{Evans2019}.
In the latter two cases, we are typically concerned with a very large (Avogadro) number of interacting particles which facilitates a statistical treatment of the problem.
Of particular interest are the one- and two-particle correlation functions of the system such as the (position-dependent) pair correlation function in classical statistical mechanics or the position and time-dependent one- and two-particle Green's functions in many-body quantum systems which describe one- and two-particle excitations.
Apart from being interesting on their own, they provide access to thermodynamic observables such as pressure, entropy or free as well as kinetic and potential energies.
The calculation of these correlation functions is, however, difficult in the presence of interactions between the particles.
For weakly interacting systems, an effective independent particle description is possible which is exploited in static mean-field theories~\cite{Dalton2022} where the interaction between the particles is replaced by a self-consistent field.
At stronger coupling this procedure yields increasingly unreliable results since the interaction between the particles must be taken into account explicitly.
As there is no exact solution to this problem for more than two particles one has to consider approximations.
For classical systems, the Ornstein-Zernicke equation~\cite{Ornstein1914} together with  specific closure relations~\cite{Hansen2013} can be exploited while quantum mechanical Green's functions can be calculated by Feynman diagrammatic perturbation theory~\cite{Abrikosov1975}. 
Unfortunately, the correlation functions obtained in this way lead to thermodynamic inconsistencies.
In classical systems thermodynamic observables such as pressure or free energy can be obtained from the pair correlation functions in different ways.
For the exact solution, all results of course coincide, however, an approximate pair correlation function typically provides different results depending on the route which is exploited for the determination of thermodynamic variables~\cite{Tsednee2019}.
Such thermodynamic inconsistencies can be also observed in the quantum case where potential and kinetic energies differ~\cite{vanLoon2016,Krien2017} when they are calculated from one- and two-particle Green's functions respectively. 
It is obvious that these discrepancies limit the predictive power of theoretical calculations.

A good example for such thermodynamics inconsistencies is the dynamical mean-field theory (DMFT)~\cite{Metzner1989,Georges1996}.
For a (finite dimensional) lattice model with purely local interactions between electrons at the same lattice site, such as the Hubbard Hamiltonian, DMFT approximates the irreducible part of the one- and two-particle Green's functions (i.e.,~the electronic self-energy $\Sigma$ and the vertex $\Gamma_r$ irreducible in the scattering channel $r$) by summing up all purely local Feynman diagrams for these correlation functions.
In this way, all purely local correlation effects in the system are captured exactly while nonlocal correlations are described on a mean-field level.
Since DMFT is a conserving theory it satisfies all conservation laws of the system (except for momentum conservation~\cite{Hettler2000}) which guarantees consistent results for the kinetic energy at the one- and the two-particle level.\footnote{Note that for a lattice system the consistency of the kinetic energy between the one- and the two-particle levels is a consequences of the $f$-sum rule which originates from particle number conservation (rather than from momentum conservation). For details see, e.g., Refs.~\cite{Vilk1997,Rohringer2016}}
However, the potential energy obtained via the one-particle self-energy $\Sigma$ differs from the one calculated by the two-particle vertex $\Gamma_r$ and the nonlocal DMFT bubble susceptibility.

A similar situation is often observed for the diagrammatic extensions of DMFT~\cite{Rohringer2018a,Toschi2007a,Rubtsov2008,Rubtsov2012,Rohringer2013,Taranto2014,Ayral2015,Ayral2016,Ayral2016a} which include nonlocal correlations beyond the local ones of DMFT by a Feynman diagrammatic expansion around the DMFT starting point.
Which of the sum rules and conservation laws are violated in this case depends of course on the actual choice of Feynman diagrams.
For example, the dynamical vertex approximation (D$\Gamma$A)~\cite{Toschi2007a,Valli2015,Eckhardt2020,Krien2020,Krien2020a,Astleithner2020} and the closely related QUADRILEX~\cite{Ayral2016} approach are based on the parquet formalism~\cite{Diatlov1957,Bickers2004,Yang2009,Tam2013,Astretsov2020} using the fully irreducible vertex of DMFT~\cite{Rohringer2011,Thunstr2018}.
The parquet equations lead to a fulfillment of the Pauli principle and all sum rules depending on the EoM, specifically guaranteeing consistency of the potential energy between the one- and two-particle levels.
However, any approximation based on the parquet formalism (i.e., any approximate choice of the fully irreducible vertex) lead to the two-particle correlation functions not fulfilling continuity equations and the related Ward identities~\cite{Smith1992,Kugler2018b,Chalupa2022}. 
Consequently, also derived conservation laws such as the $f$-sum rule, which follows from the continuity equation for the particle density, are generally violated in these situation. 
This implies different results for the kinetic energy when it is obtained from one- and two-particle correlation functions.

Another limitation of the parquet formalism is its very high numerical cost even for single-orbital models, making its extension to multi-orbital systems unrealistic in the foreseeable future. 
Hence, other routes have been pursued to achieve consistent results for the potential energies. 
Within the dual boson (DB) method~\cite{Stepanov2016,Peters2019}, a purely local reference system (analogous to DMFT) with an effective frequency dependent interaction is introduced. 
The latter is determined by the condition that the local parts of the lattice charge and spin susceptibilities are equal to the corresponding ones of the local reference system. 
While this approach certainly improves the consistency between the one- and the two-particle levels, it has been recently shown that an additional term in the calculation of the potential energy from the one-particle correlation function emerges due to the frequency dependence of the interaction~\cite{Krien2017}, which again destroys the consistency.
Moreover, the DB approach requires the repeated solution of an effective Anderson impurity model (AIM) with a frequency dependent interaction making it numerically challenging.

In this work, we will consider the consistency of the potential energy within another diagrammatic extension of DMFT, the ladder version of the dynamical vertex approximation (D$\Gamma$A)~\cite{Katanin2009,Rohringer2011,Schafer2015,Schafer2017,Schaefer2019}.
Within this approach, the sum rule for the up-up susceptibility (which corresponds to the Pauli principle) has been already restored by means of a so-called $\lambda$-correction parameter in the spin channel.\footnote{Let us also mention a recent development, where a higher degree of self-consistency has been achieved in the ladder D$\Gamma$A by considering a self-consistent feedback of the one-particle self-energy on the calculation of the two-particle susceptibilities as proposed in Ref.~\onlinecite{Kaufmann2021}}
More specifically, the correlation length of the D$\Gamma$A spin susceptibility is renormalized with a constant parameter $\lmagn$ determined by the above mentioned sum rule. 
Here we will extend this idea~\cite{Rohringer2016} by correcting also the charge susceptibility by a second parameter $\ldens$.
Both parameters are then simultaneously fixed by the sum rule for the up-up susceptibility and the requirement that the potential energies at the one- and the two-particle level should be equivalent.
This idea shares some similarities with the two-particle self-consistent approach (TPSC)~\cite{Vilk1994,Vilk1997} where the fulfillment of two-particle consistency is achieved by considering (different) effective Hubbard interaction parameters in the charge and spin channel. 
However, the latter approach is restricted to the weak coupling regime, while the DMFT input in our improved version of ladder D$\Gamma$A makes the method applicable to the entire range of the coupling parameter.
With this method, we achieve an improved description of the phase diagram in the weak-to-intermediate coupling regime and consistent results for the potential energy in the entire parameter space.
Finally, we also restore the correct hierarchy of the kinetic energy between D$\Gamma$A and DMFT in the weak-to-intermediate coupling regime.

The paper is organized as follows.
In Sec.~\ref{sec:Method}, we discuss the general formalism of the ladder D$\Gamma$A approach and introduce our new method. 
In Sec.~\ref{sec:results}, we present our results for the three-dimensional half-filled Hubbard model on a simple cubic lattice and discuss the impact of the extended $\lambda$-correction scheme on charge and spin susceptibilities, phase diagram, electronic self-energies as well as on the potential and kinetic energies.
In Sec.~\ref{sec:conclusions} we conclude our work.

\section{Method}\label{sec:Method}

In this study we will consider the single-band Hubbard model,
\begin{equation}\label{equ:hubbard_h}
    \hat{\mathcal{H}} = - t \sum_{\langle ij \rangle \sigma} \hat{c}^\dagger_{i\sigma} \hat{c}_{j\sigma} + U \sum_i \hat{n}_{i\uparrow} \hat{n}_{i\downarrow}
\end{equation}
with hopping amplitude $t$ between nearest neighbors and effective Hubbard interaction $U$ between particles at the same lattice site.
$\hat{c}_{i\sigma}^{(\dagger)}$ annihilates (creates) an electron with spin $\sigma$ at lattice site $\mathbf{R}_i$ and $\hat{n}_{i\sigma}=\hat{c}^\dagger_{i\sigma}\hat{c}_{i\sigma}$ is the local density.
We will restrict ourselves to the half-filled ($n=1$) three-dimensional simple cubic lattice with nearest neighbor hopping.
The energy scale will be fixed to $D = 2 \sqrt{6}t$ which corresponds to twice the standard deviation of the noninteracting density of states.
Furthermore, we will use $\nu=(2n+1)\frac{\pi}{\beta}$, $n\in\mathds{Z}$, to indicate fermionic and $\omega=2m\frac{\pi}{\beta}$, $m\in\mathds{Z}$, for bosonic Matsubara frequencies.
$\beta=1/T$ denotes the inverse temperature of the system. 
Lastly, the factor of $\frac{1}{\beta}$ for Matsubara sums is omitted, i.e.~$\sum_\nu \coloneqq \frac{1}{\beta} \sum_{n=-\infty}^{\infty}$ and integrals over the momentum vectors $\bk$ or $\bq$ over the Brillouin zone (BZ) are written as sums $\sum_\bk \coloneqq \frac{1}{V_\mathrm{BZ}} \int_\mathrm{BZ} \diff \bk$, where $V_\text{BZ}$ is the volume of the BZ.

\subsection{Ladder \texorpdfstring{D$\Gamma$A}{DGA} formalism}
The method employed in this work is based on D$\Gamma$A~\cite{Toschi2007a} in its ladder approximation~\cite{Katanin2009,Rohringer2016}.
D$\Gamma$A is a natural generalization of DMFT in the following sense:
DMFT assumes the one-particle irreducible (1PI) one-particle vertex, the electronic self-energy $\Sigma^\nu$, to be purely local, i.e.,~$\bk$ independent.
D$\Gamma$A raises this concept to the two-particle level and assumes the $2$PI  two-particle vertex $\Lambda^{\nu\nu'\omega}_{\sigma\sigma'}$ to be local. 
This is a systematic approximation in the sense that the theory becomes exact in the limit of $n \to \infty$ for local $n$PI vertices.
$\Lambda_{\sigma\sigma'}^{\nu\nu'\omega}$ introduces nonlocal correlation effects on top of the local ones of DMFT via a momentum dependent self-energy $\Sigma_\mathbf{k}^\nu$ which is obtained from the equation of motion
\vspace*{3mm}
\begin{align}
    \Sigma^\nu_\bk  = \frac{Un}{2} -  & \, U \sum\displaylimits_{\mathclap{\nu'\omega\bk'\bq}}  F^{\nu\nu'\omega}_{\uparrow\downarrow, \bk\bk'\bq}  G^{\nu'}_{\bk'}  G^{\nu'+\omega}_{\bk'+\bq}G^{\nu+\omega}_{\bk + \bq}. \label{equ:eom} 
\end{align}
\vspace*{3mm}\\
The full vertex $F^{\nu\nu'\omega}_{\uparrow\downarrow, \bk\bk'\bq}$ is calculated from $\Lambda^{\nu\nu'\omega}_{\sigma\sigma'}$ through the Bethe-Salpeter and parquet equations~\cite{Diatlov1957,Tam2013}. 
The former connects $\Lambda^{\nu\nu'\omega}_{\sigma\sigma'}$ with $F^{\nu\nu'\omega}_{\uparrow\downarrow, \bk\bk'\bq}$ in three different ways, corresponding to fluctuations in the charge, spin and particle-particle channels. 
The latter takes into account the mutual renormalization effects between these three different channels. 
However, this approach is numerically very expensive.
The ladder approximation therefore omits the self consistency and calculates the full vertex only once via a one-shot Bethe-Salpeter equation in the relevant scattering channels. 
For the repulsive particle-hole symmetric Hubbard model, these are the charge (density, $r=\text{d}$) and the spin (magnetic, $r=\text{m}$) channel while particle-particle fluctuation are typically strongly suppressed and sufficiently well captured at the local level by DMFT. 
Moreover, also the Green's functions appearing in Eq.~(\ref{equ:eom}) remain on the DMFT level, contrary to self consistent methods like parquet.
Unfortunately, the ladder approximation violates the two-particle self-consistency. 
This leads to (i) a spurious $1/i\nu$ asymptotic behavior of the self-energy~\cite{Rohringer2016} and (ii) ambiguous results for the potential energy. 
To overcome these problems it is necessary to rewrite Eq.~(\ref{equ:eom}) in terms of physical susceptibilities. 
To this end we introduce the bare, generalized and physical susceptibilities as well as the triangular vertex which are defined as follows (the upper sign corresponds to the spin, the lower to the charge channel):
\vspace*{3mm}
\begin{align}
    \chi^{\nu\nu'\omega}_{0,\bq} & = - \beta \delta_{\nu\nu'} \sum_{\bk} G^{\nu}_\bk G^{\nu+\omega}_{\bk + \bq}, \label{equ:chi_0_def}\\
    \chi^{\nu\nu'\omega}_{r,\bq} & = \chi^{\nu\nu'\omega}_{0,\bq} - \sum_{\nu_1\nu_2} \chi^{\nu\nu_1\omega}_{0,\bq}\Gamma^{\nu_1\nu_2\omega}_r \chi^{\nu_2\nu'\omega}_{r,\bq}, \label{equ:chi_gen_def}\\
    \chi^\omega_{r,\bq} & = \sum_{\nu\nu'} \chi^{\nu\nu'\omega}_{r,\bq}, \label{equ:chi_phys_def}\\
    \gamma^{\nu\omega}_{r,\bq} & = \sum_{\nu'} \left( \chi^{\nu\nu'\omega}_{0,\bq} (1 \pm U \chi^{\omega}_{r,\bq} ) \right)^{-1} \chi^{\nu\nu'\omega}_{r,\bq}. \label{equ:gamma_def}
\end{align}
\vspace*{3mm}\\
$G_\mathbf{k}^\nu=[i\nu+\mu-\varepsilon_\mathbf{k}-\Sigma^\nu ]^{-1}$ is the DMFT lattice Green's function and $\Sigma^\nu$ the local DMFT self-energy. 
$\Gamma_r^{\nu\nu'\omega}$ denotes the local vertex which is irreducible in channel $r$.
These quantities allow us to reformulate the equation of motion~\cite{Rohringer2011,Katanin2009,Rohringer2016}:
\begin{align}
    \Sigma^{\bl,\nu}_{\bk} = \frac{Un}{2} -  U & \sum\displaylimits_{\omega\bq} \Big[ 1 + \frac{1}{2} \gamma^{\nu\omega}_{\text{d},\bq} (1 - U \chi^{\lambda_d,\omega}_{\text{d},\bq}) \nonumber\\
    & - \frac{3}{2}\gamma^{\nu\omega}_{\text{m},\bq}(1 + U \chi^{\lambda_m,\omega}_{\text{m},\bq})\, -  \sum\displaylimits_{\nu'} \chi^{\nu'\omega}_{0,\bq} F_{\mathrm{m}}^{\nu\nu'\omega} \Big] G^{\nu+\omega}_{\bk + \bq}, \label{equ:eom_2} 
\end{align}
where $F_{\text{m}}^{\nu\nu'\omega}$ is the local full vertex of DMFT.
A detailed discussion of the method is given in Ref.~\cite{Rohringer2016} [c.f., Sec.~I.A., Eq.~(2) to Eq.~(10) in this reference].

With these definitions we can formulate the above discussed consistency relations for the Pauli principle and the potential energy as:
\begin{subequations}
\label{subequ:lambdacondition}
\begin{align}
        \frac{1}{2}\sum_{\omega\bq}\left(\chi^{\ldens,\omega}_{\mathrm{d},\bq} 
            + \chi^{\lmagn, \omega}_{\mathrm{m},\bq}\right)
%    =\sum_{\omega\bq} \chi^{\bl,\omega}_{\uparrow\uparrow, \bq}
    & \overset{!}{=} 
        \frac{n}{2} \left( 1 - \frac{n}{2} \right) \label{equ:lambdacondition1}\\
%    \frac{U}{2}\sum_{\omega\bq}\left(\chi^{\ldens,\omega}_{\mathrm{d},\bq}
%        - \chi^{\lmagn, \omega}_{\mathrm{m},\bq}\right)
%        + U\frac{n^2}{4}\nonumber\\
%    =\underbrace{U \sum_{\omega\bq} \chi^{\bl,\omega}_{\uparrow\downarrow,\bq} 
%        + U\frac{n^2}{4}}_{E^{(2)}_\text{pot}}
    \underbrace{\frac{U}{2}\sum_{\omega\bq}\left(\chi^{\ldens,\omega}_{\mathrm{d},\bq}
        - \chi^{\lmagn, \omega}_{\mathrm{m},\bq}\right)
        + U\frac{n^2}{4}}_{E^{(2)}_\text{pot}}
    & \overset{!}{=} 
        \underbrace{\sum_{\nu\bk} G^{\bl,\nu}_\bk \Sigma^{\bl,\nu}_{\bk}}_{E^{(1)}_\text{pot}}, \label{equ:lambdacondition2}
\end{align}
\end{subequations}
where we introduced [here and also in Eq.~(\ref{equ:eom_2})] the free parameters $\ldens$ and $\lmagn$ to fulfill these consistency relations. Note that Eq.~(\ref{equ:gamma_def}) is not affected by any $\lambda$-correction as the triangular vertex already features the correct high-frequency asymptotic behavior. 
More details are given in Sec.~V of Ref.~\onlinecite{Rohringer2016}.
The $\lambda$'s enter into the formalism via a renormalization of the physical susceptibilities in the spirit of the Moriya theory of itinerant magnetism~\cite{Moriya1985} as follows
\begin{equation}
    \chi^{\lambda_r}_r = \left( \frac{1}{\chi_r} + \lambda_r \right)^{-1}. \label{equ:lambdacorr}
\end{equation}

A previous version of ladder D$\Gamma$A\cite{Katanin2009} has already exploited a simpler type of this idea where the sum rule in Eq.~(\ref{equ:lambdacondition1}) has been enforced by considering a $\lambda$-correction {\em only} in the spin channel (i.e.,~$\ldens=0$). 
In our new approach we achieve a higher degree of consistency by overcoming the ambiguity in the determination of the potential energy with a corresponding renormalization in the charge channel.

\section{Results}
\label{sec:results}

In the following we present results for the charge (density) and spin (magnetic) susceptibilities and the related magnetic phase diagram, the self-energies as well as the potential and kinetic energies, which are obtained by our method. 
We focus particularly on the mutual renormalization effects between charge and spin fluctuations which are introduced by the consistency relations in Eqs.~(\ref{subequ:lambdacondition}). 
Moreover, we compare our findings with previous ladder D$\Gamma$A calculations~\cite{Katanin2009,Rohringer2011,Rohringer2016,Schafer2017}, where only the spin channel has been renormalized, as well as to other diagrammatic and numerical techniques.

To more concisely distinguish the different methods, we use the following notation: The previous version of D$\Gamma$A will be denoted with \lamm. 
The index ``m'' indicates that only the magnetic susceptibility is renormalized by a parameter $\lmagn \ne 0$ while $\ldens = 0$. 
\lamdm refers to our new approach where both the charge and spin susceptibility are corrected by renormalization constants $\ldens$, $\lmagn \ne 0$.

The local DMFT self-energy $\Sigma^\nu$ and vertex functions $\Gamma_{r}^{\nu\nu'\omega}$ have been obtained from an exact diagonalization (ED) impurity solver using four bath sites.
While the applicability of ED is certainly limited by the necessity of fitting the hybridization function to a finite bath, it does not suffer from any statistical noise which typically arises in quantum Monte Carlo calculations.
The latter particularly affects the two-particle vertex $\Gamma_{r}^{\nu\nu'\omega}$ which is obtained from a matrix inversion in the space of the fermionic Matsubara frequencies $\nu$ and $\nu'$. 
For the development of the new method we deemed statistical fluctuations of the input data as problematic, since the effect of specific features in the approach and the statistical error on the results cannot easily be disentangled. 
Such problems are indeed absent in ED which is in any case expected to provide reliable results at the rather high temperatures above the $\TN$ of D$\Gamma$A. 
We have nevertheless checked our numerical findings for a broad range of points in the $U$ versus $T$ phase diagram with continuous time quantum Monte Carlo (CTQMC) calculations in the hybridization expansion implementation using the w2dynamics package~\cite{Wallerberger2019}.

\subsection{Physical interpretation and determination of \texorpdfstring{$\ldens$ and $\lmagn$}{parameters}}\label{sec:determinationlambda}

\begin{figure}[t!]
    \centering
    \includegraphics[width=\linewidth]{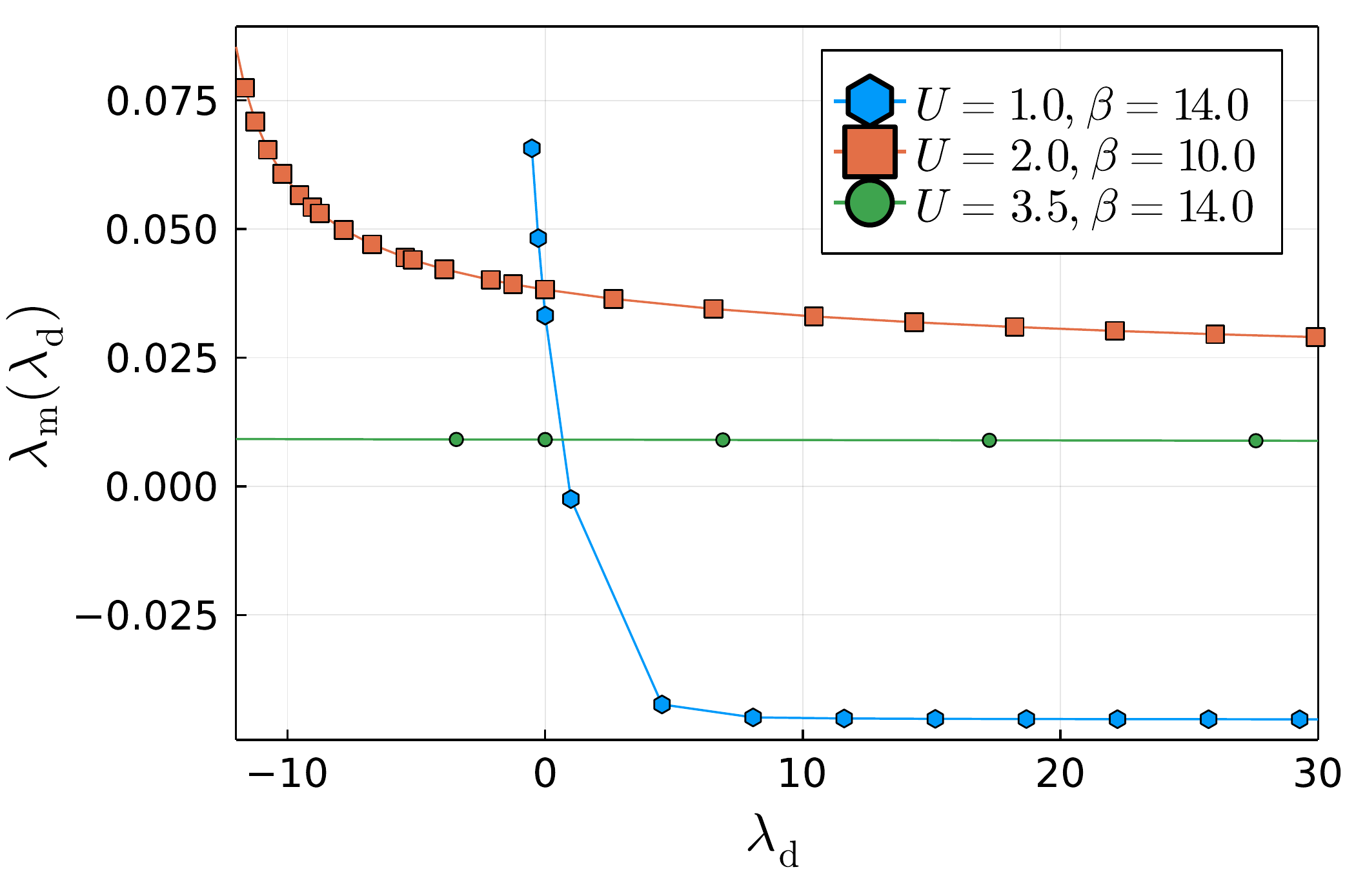}
    \caption{$\lmagn$ as a function of $\ldens$ for three different values of $U$ at $\beta=10$ and $\beta=14$. The divergence of $\lmagn(\ldens)$ indicates the largest pole of $\chi_{\text{d},\bq}^{\ldens,\omega}$ as a function of $\ldens$ [see Eq.~(\ref{equ:lambdacorr})] after which this susceptibility would become negative making all solutions with smaller $\ldens$ unphysical.}
    \label{fig:lambda_sp_of_ch_res}
\end{figure}

\begin{figure}[t!]
    \centering
    \includegraphics[width=\linewidth]{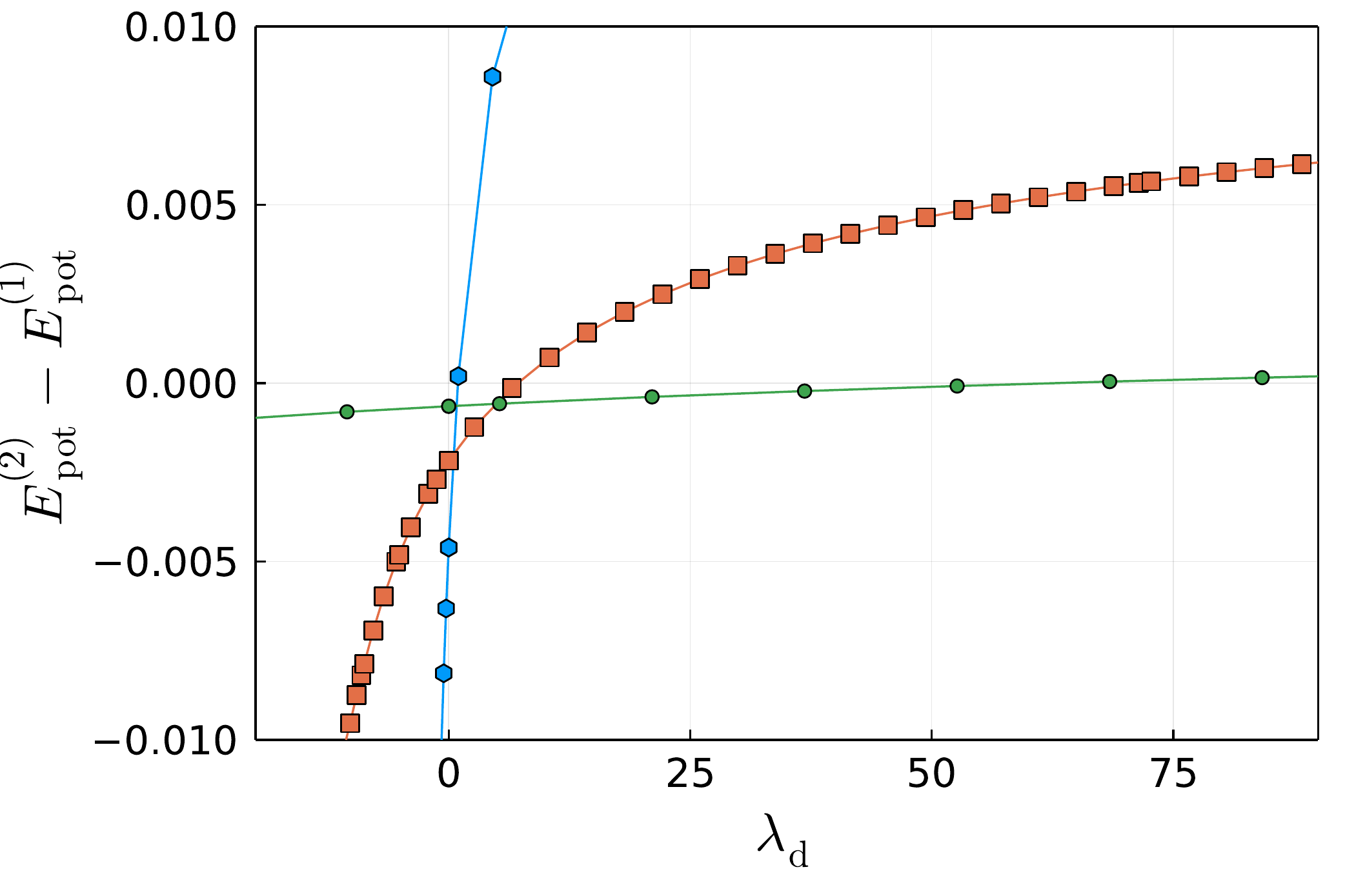}
    \caption{Difference between the potential energies obtained from one- and two-particle correlation functions as on the right and left-hand sides of Eq.~(\ref{equ:lambdacondition2}) respectively as a function of $\ldens$ where $\lmagn$ is obtained (for a given $\ldens$) from Eq.~(\ref{equ:lambdacondition1}). Results at three different values of $U$ at $\beta=10$ and $\beta=14$ are shown, corresponding to weak, intermediate and strong coupling. The crossing of the lines with the $x$-axis indicates a solution for the set of consistency Eqs.~(\ref{subequ:lambdacondition}).}
    \label{fig:c2_curved_3U}
\end{figure}

%The $\lambda$ parameters introduced in Eq.~\ref{equ:lambdacorr} serve as an effective renormalization of the correlation length.  As discussed above, they encapsulate two different effects in this way: (1) Extrapolation from the one-shot approximation of the ladder D$\Gamma$A approach. (2) Introduction of cross-effects between the channels through Eq.~\ref{equ:lambdacondition2}.

%It is therefore worth discussing some details here. Additionally, several numerical challenges arise when trying to obtain a solution with usual, straight forward methods one possible solution method presents a clearer picture of the rebalancing effect between the channels. Details for a numerically more efficient algorithm are discussed in Sec.~\ref{sec:root_finding}.

Considering an Ornstein-Zernike form for the physical charge and spin susceptibilities 
\begin{equation}
\label{equ:ornstein}
\chi^{\omega_0}_{r,\bq} \sim\frac{1}{\mathbf{q}^2-\xi_r^{-2}}, 
\end{equation}
it is obvious that the $\lambda$-corrections introduced in Eq.~(\ref{equ:lambdacorr}) correspond to a renormalization of the correlation length $(\xi_r)^{-2}\rightarrow(\xi_{r}^{\lambda_r})^{-2}=\xi_{r}^{-2}+\lambda_r$ or, after rewriting, $\xi_r\rightarrow\xi_{r}^{\lambda_r}=\frac{\xi_r}{\sqrt{1+\lambda_r\xi_r}}$.
Let us remark that, from field theoretical perspective, $\xi_r^{-2}$ corresponds to the mass of the propagator of the corresponding charge and spin fluctuations and, hence, $\lambda_r$ can be also interpreted as a mass renormalization. 
The actual values of the parameters $\ldens$ and $\lmagn$ are determined be the consistency relations Eq.~(\ref{equ:lambdacondition1}) for the Pauli principle and Eq.~(\ref{equ:lambdacondition2}) for the potential energy, respectively. 
A numerically efficient algorithm to determine $\ldens$ and $\lmagn$ from these equations is discussed in Appendix~\ref{sec:root_finding}. 
Here, instead, we present a solution method which better highlights the physical content of our approach. 
This method consists in two steps:

First, we use only sum rule Eq.~(\ref{equ:lambdacondition1}), corresponding to the Pauli principle, to calculate $\lmagn$ for given values of $\ldens$.
In this way we obtain a function $\lmagn(\ldens)$ which is depicted in Fig.~\ref{fig:lambda_sp_of_ch_res} for three different values of $U$ at $\beta=10$ and $\beta=14$.
The values for the temperature are chosen such that the distance to the phase transition is similar for the three values of $U$ (c.f. Sec.~\ref{sec:phasediagram}).
The allowed range of values for $\ldens$ and $\lmagn$ is determined by the condition that both the density and the magnetic susceptibilities $\cdensF$ and $\cmagnF$ must be real and positive for all frequencies $\omega$ and momenta $\mathbf{q}$ (see also Appendix~\ref{sec:root_finding} for a discussion of the physically relevant interval). 
We observe that $\lmagn$ is monotonously decreasing with increasing $\ldens$.
This behavior can be directly understood from Eq.~(\ref{equ:lambdacondition1}): a larger value of $\ldens$ corresponds to a smaller $\cdensF$.
Hence, in order to fulfill this sum rule the decrease of $\cdensF$ must be compensated by a corresponding increase of $\cmagnF$ which is achieved by lowering the value of $\lmagn$.
Therefore Eq.~(\ref{equ:lambdacondition1}) provides in a simplified way the {\em mutual} renormalization of the charge and spin susceptibilities as it is usually achieved only by a full parquet treatment~\cite{Bickers2004,Valli2015,Krien2020} of the problem. 

\begin{figure}
    \centering
    \includegraphics[width=\linewidth]{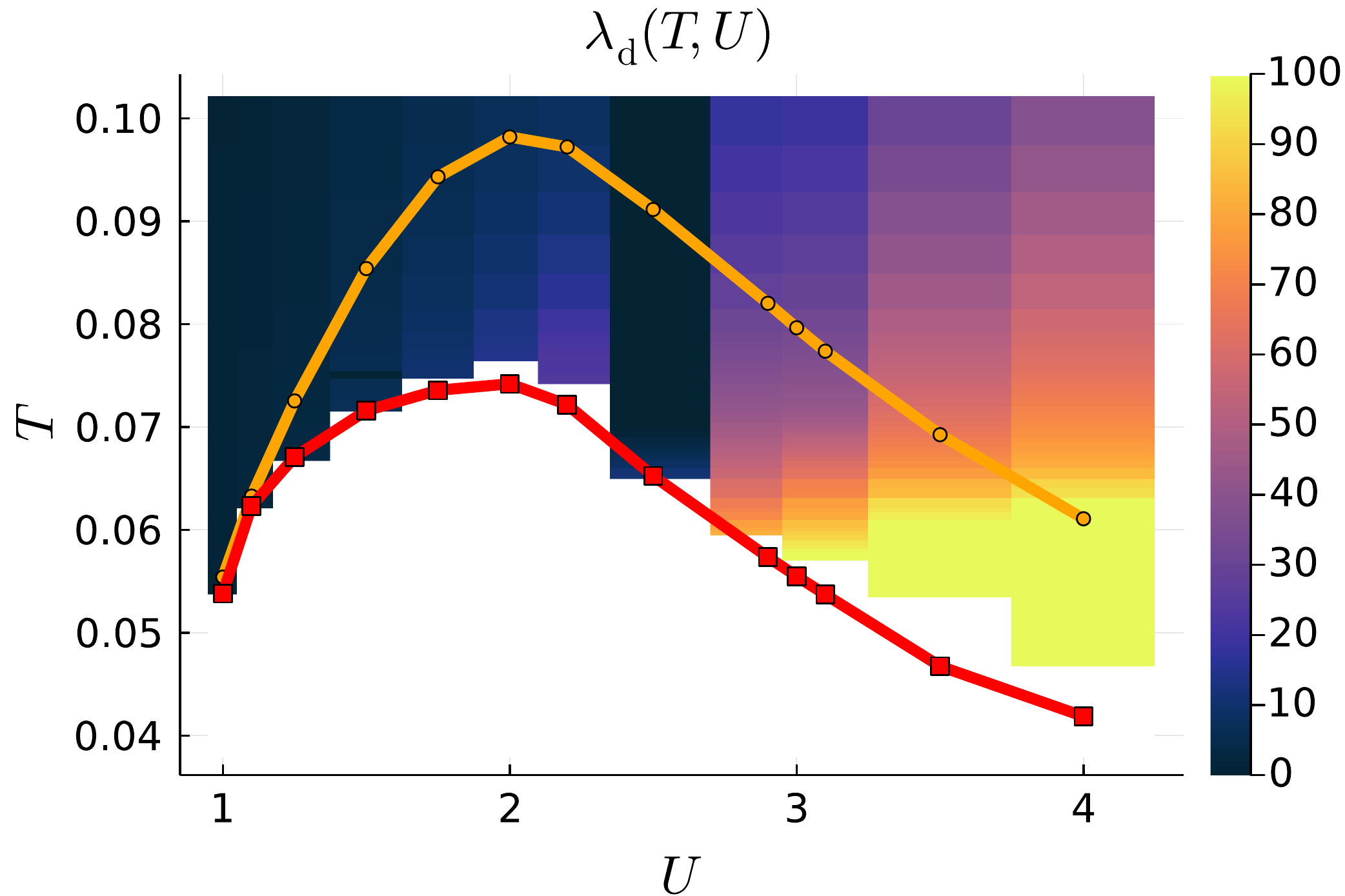}
    \caption{Heat map of $\ldens$ as a function of $U$ and $T$ from weak to strong coupling. The orange line indicates the N\'eel temperature of DMFT while the red line corresponds to the $\TN$ of \lamdm.}
    \label{fig:lambdach_grid}
\end{figure}

To determine the value of $\ldens$ we have to consider Eq.~(\ref{equ:lambdacondition2}) which corresponds to the consistency of the potential energies between the one- and the two-particle level.
In Fig.~\ref{fig:c2_curved_3U} we show the difference between the left and the right-hand side of this equation as a function of $\ldens$ for three different values of $U$ at $\beta=10$ and $\beta=14$ where $\lambda_\text{m}=\lambda_\text{m}(\lambda_\text{d})$. 
%How to obtain the corresponding $\lmagn(\ldens)$ from Eq.~(\ref{equ:lambdacondition1}) is discussed above.
The value of $\ldens$ at which the curve crosses zero corresponds to a solution of the consistency equation for the potential energy Eq.~(\ref{equ:lambdacondition2}).
For each of the considered values of $U$ we find such a crossing for positive values of $\ldens$. 
Moreover, we observe that the slope of the lines decreases with increasing interaction strength.
This behavior can be attributed to the overall magnitude of charge fluctuations in the respective parameter regime. 
At weak coupling ($U=1$), $\cdensF$ is still comparatively large and, hence, its inverse is small.
Correcting a small value by $\ldens$ and inverting again [see Eq.~(\ref{equ:lambdacorr})] results in a substantial modification of $\cdensF$ and all quantities depending on it.
On the contrary, at larger values of $U$ ($U=2$ and $U=3.5$) close to or beyond the Mott metal-to-insulator transition charge fluctuations are strongly suppressed and $\cdensF$ becomes very small. 
Consequently its inverse gets very large and is only weakly affected by the addition of the parameter $\ldens$ in Eq.~(\ref{equ:lambdacorr}) explaining the overall weaker dependence of observables on $\ldens$ at strong coupling. 
This observation has also implications for the numerical determination of $\ldens$: In fact, the calculation of $\ldens$ becomes gradually more difficult upon increasing $U$ as the correction of the already strongly suppressed charge susceptibility requires an increasingly higher numerical precision.
This also implies that the solution starts to depend stronger on small changes in the DMFT input in this parameter regime which requires a particularly precise evaluation of the DMFT correlation functions $\Sigma^\nu$ and $\Gamma_{r}^{\nu\nu'\omega}$.

Let us briefly address the signs of the (real) renormalization parameters $\ldens$ and $\lmagn$. 
On general grounds we expect that DMFT overestimates nonlocal fluctuations described by $\chi^{\omega}_{r,\bq}$ as it is a mean-field theory with respect to spatial degrees of freedom. 
Hence, the $\lambda$-corrections should {\em suppress} these DMFT fluctuations by assuming positive values $\lambda_r>0$.
This assertion is indeed true in the entire parameter regime as can be seen in Figs.~\ref{fig:c2_curved_3U}-\ref{fig:lambdach_grid}. 
In Fig.~\ref{fig:c2_curved_3U}, the curve corresponding to the consistency relation for the potential energy crosses zero at positive values of $\ldens$ for all $U$ which is confirmed by the heatmap of $\ldens$ as a function of $U$ and $T$ in Fig.~\ref{fig:lambdach_grid}. 
In the latter, we indeed observe $\ldens>0$ for all values of $U$ and $\beta$ which is also true for $\lmagn=\lmagn(\ldens)$.

\subsection{Density and magnetic susceptibilities}\label{sec:susceptibilities}

\begin{figure*}[htb]
    \centering
    \subfloat[\label{fig:chisp_loc_weak}]{
         \centering
         \includegraphics[width=0.32\linewidth]{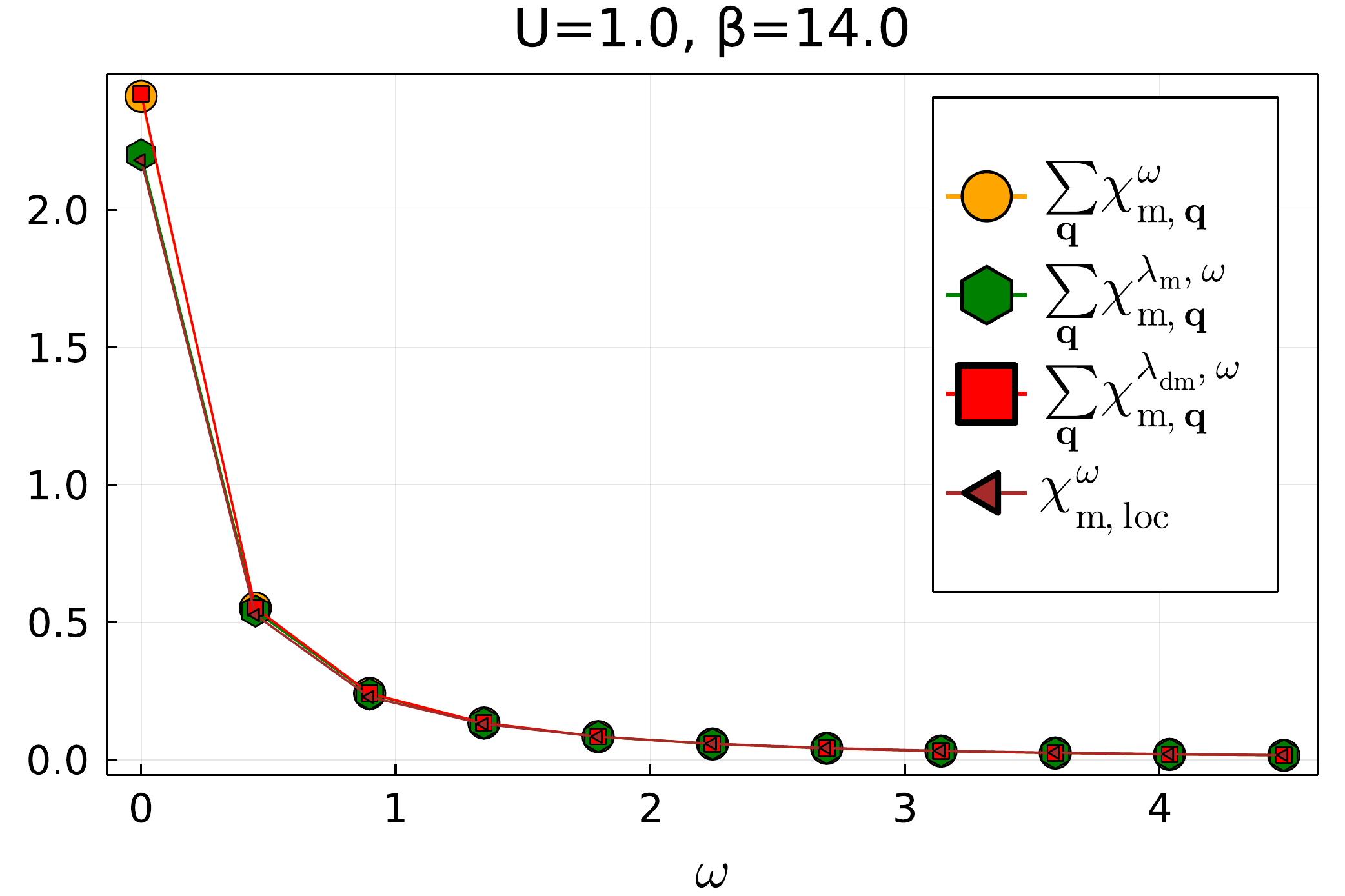}
     }
     \subfloat[\label{fig:chisp_loc_interm1k}]{
         \centering
         \includegraphics[width=0.32\linewidth]{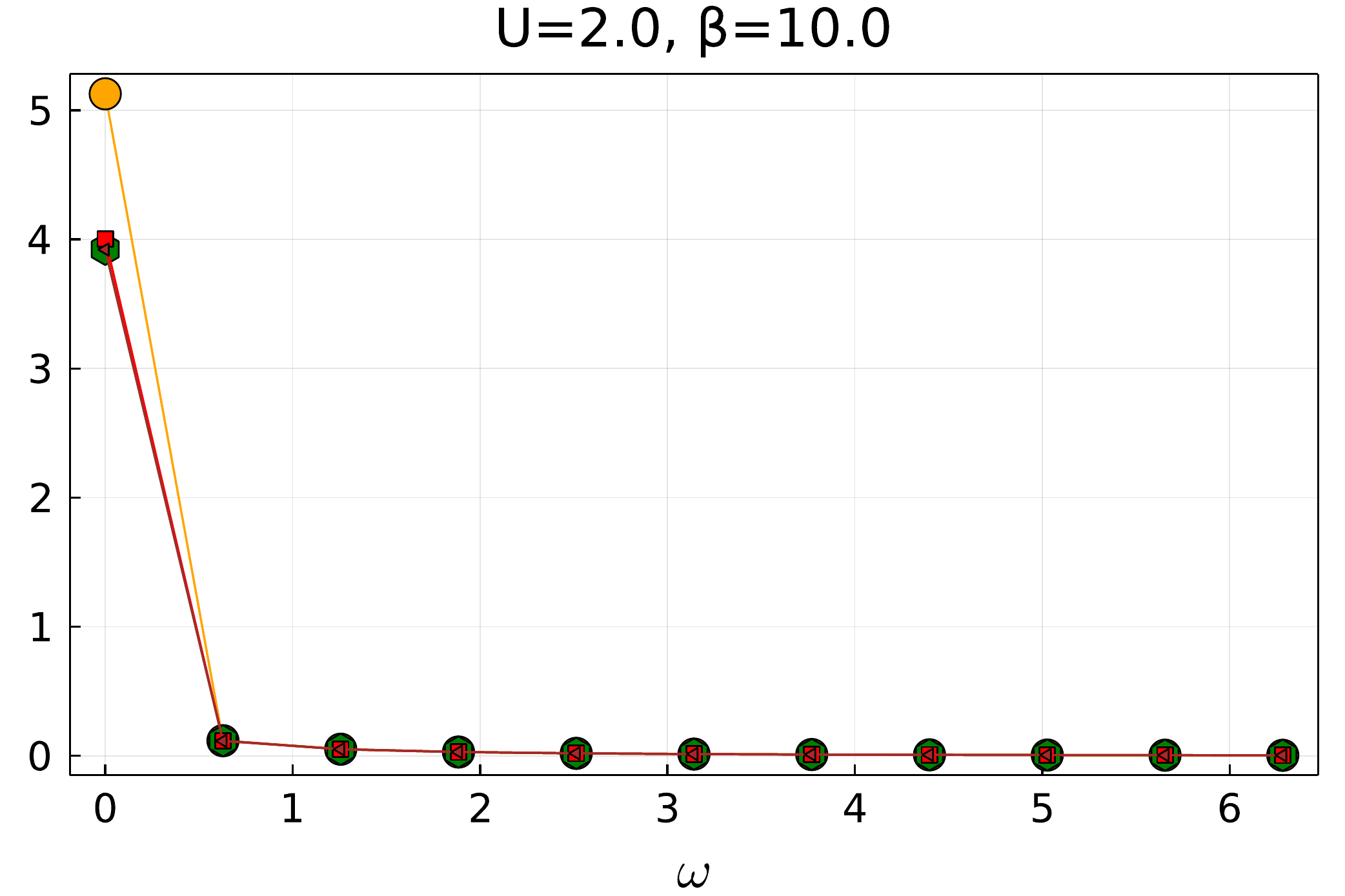}
     }
     \subfloat[\label{fig:chisp_loc_interm2}]{
         \centering
         \includegraphics[width=0.32\linewidth]{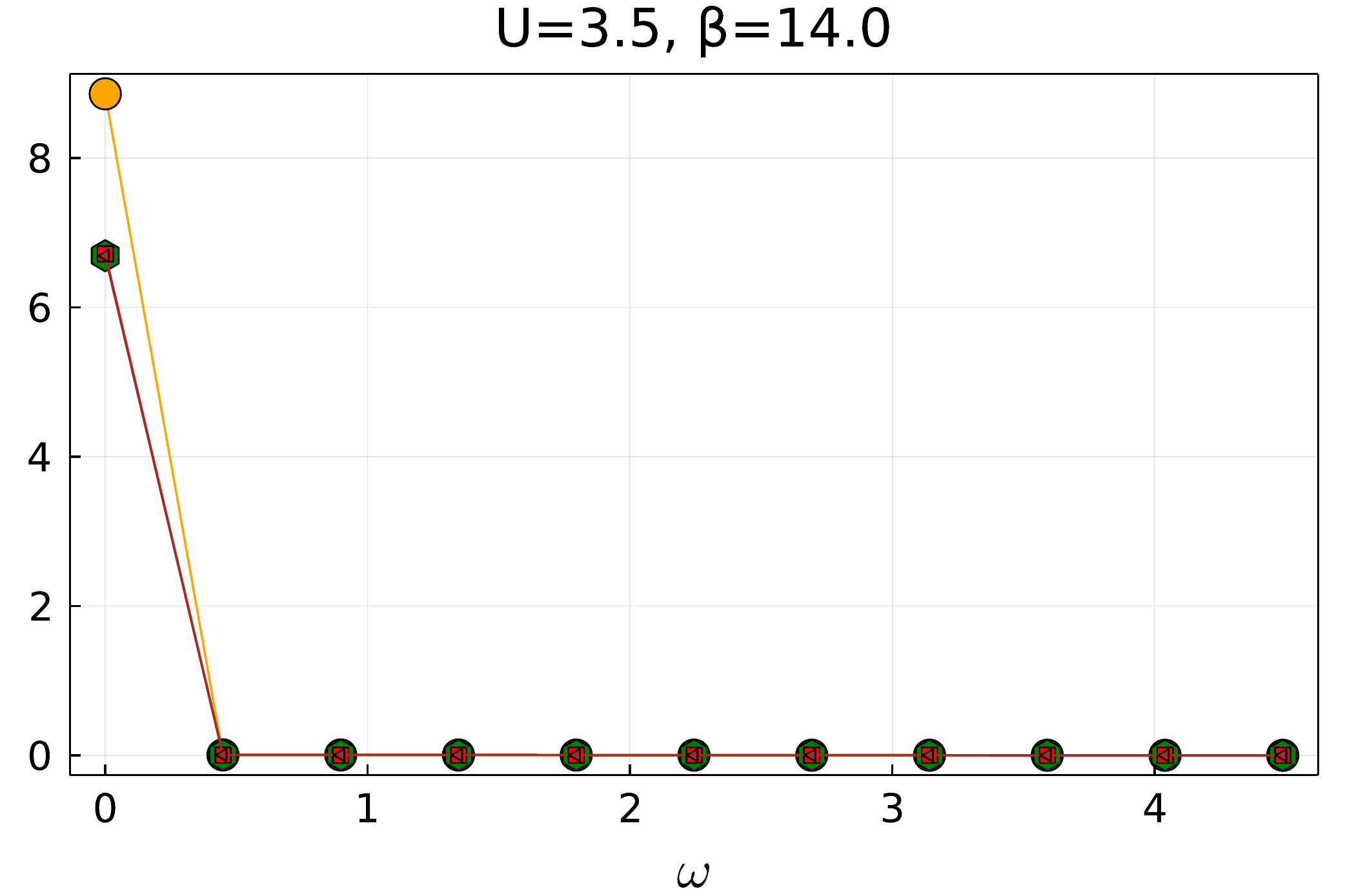}
    }
    \hfill
          \subfloat[\label{fig:chich_loc_weak}]{
         \centering
         \includegraphics[width=0.32\linewidth]{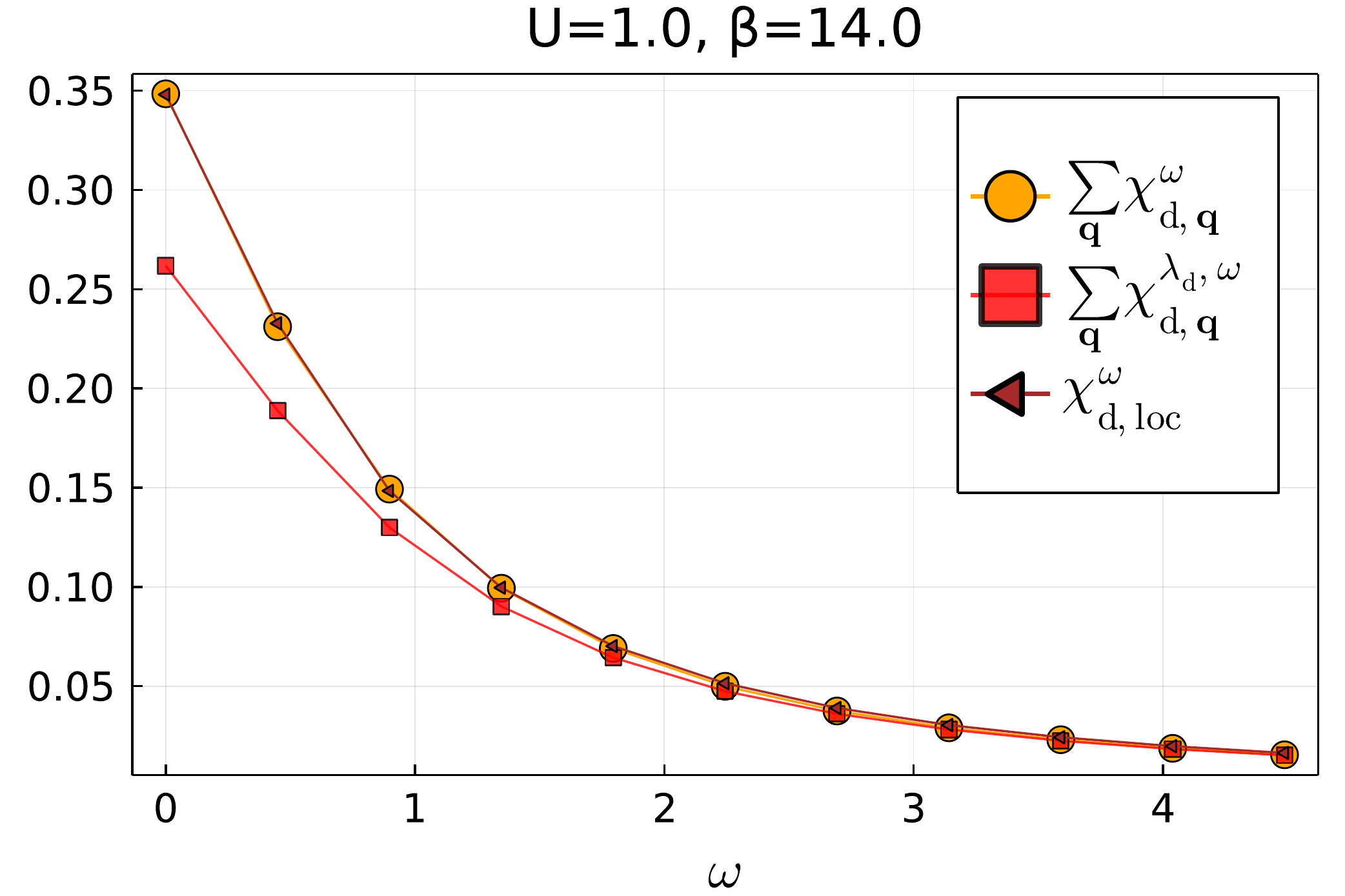}
    }
    \subfloat[\label{fig:chisp_loc_interm1}]{
        \centering
        \includegraphics[width=0.32\linewidth]{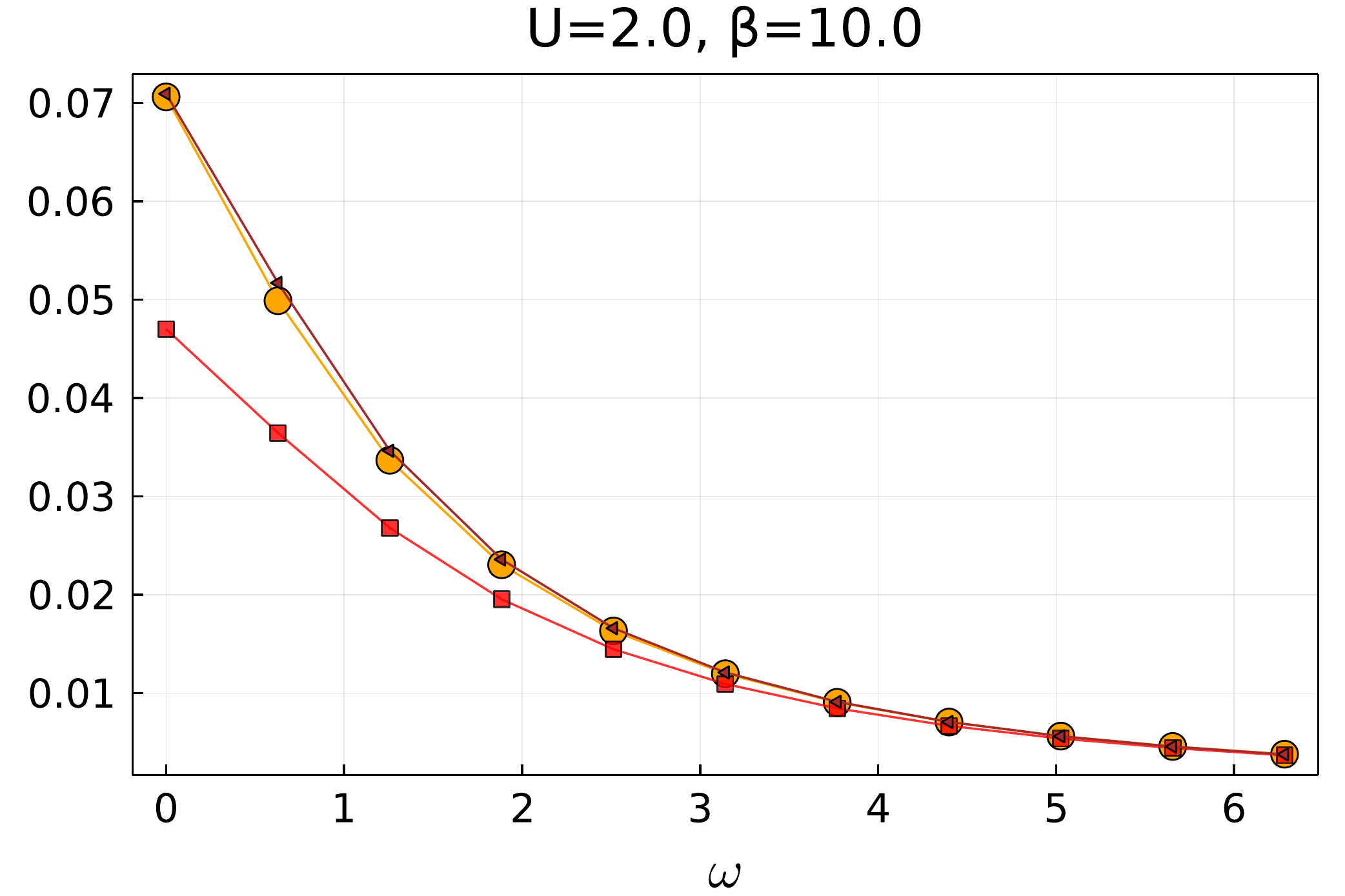}
    }
    \subfloat[\label{fig:chich_loc_interm2}]{
        \centering
        \includegraphics[width=0.32\linewidth]{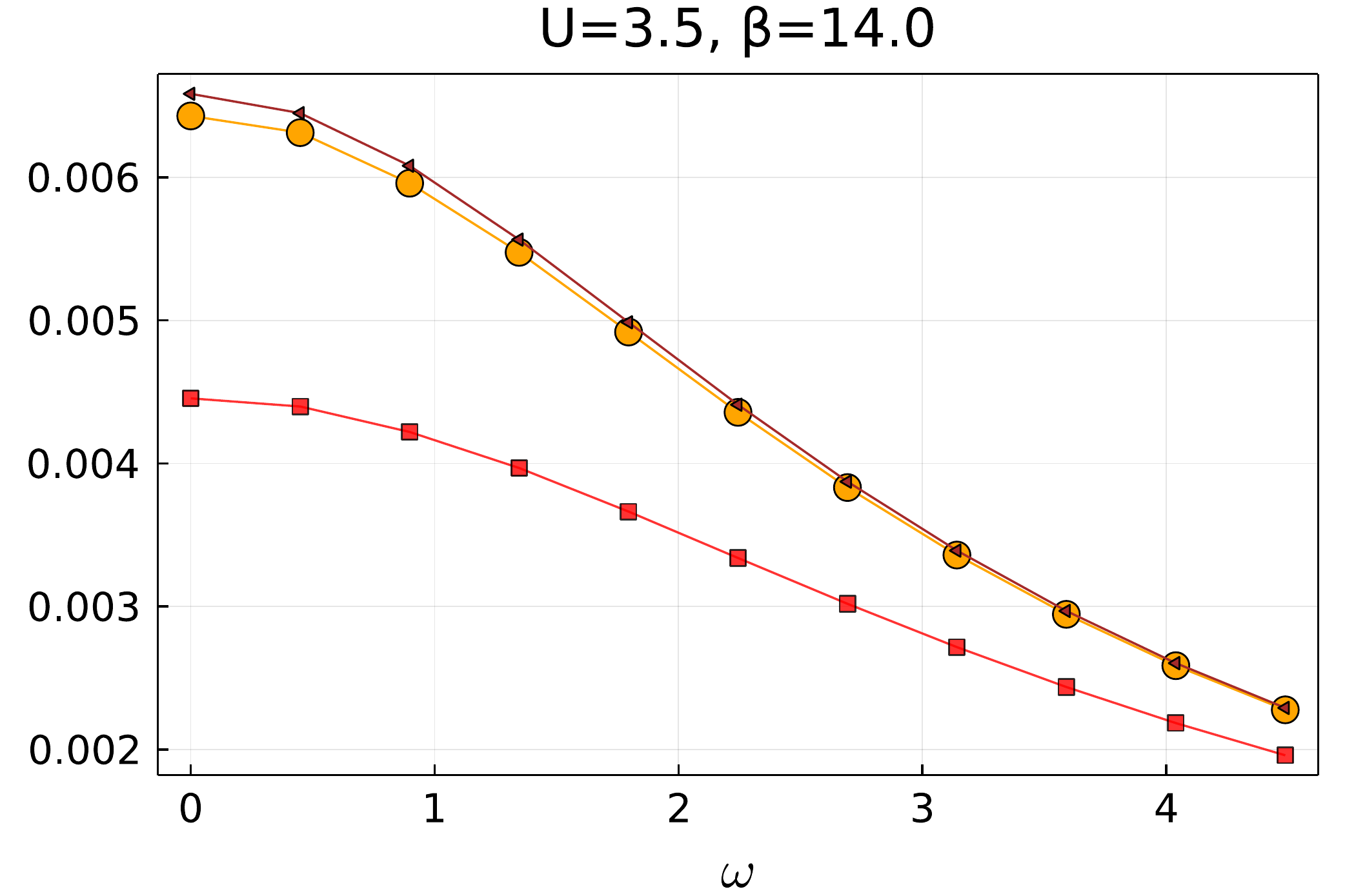}
    }
     \caption{Momentum integrated susceptibilities in spin (top row) and charge (bottom row) channel for \lamdm (red squares) versus DMFT (i.e.,~$\ldens=\lmagn=0$, orange circles) and \lamm (green hexagons). Data is presented in the weak ($U=1$), intermediate ($U=2$) and strong ($U=3.5$) coupling regime above the critical temperature of DMFT. Note that for \lamm no charge renormalization is performed making it equivalent to the DMFT solution in the lower panel. The local impurity susceptibilities of DMFT (brown triangles) are shown for comparison.}
    \label{fig:suscomega}
\end{figure*}

In this section, we discuss the lattice susceptibilities $\chi^\omega_{r,\bq}$ in the charge ($r=\dens$) and spin ($r=\magn$) channels. 
These observables are interesting on their own as they are subject to the renormalization procedure discussed in the previous section.
Moreover, they determine the antiferromagnetic phase transition and transfer the effects of charge and spin renormalization to the electronic self-energy as well as to the potential (and kinetic) energies via Eqs.~(\ref{equ:eom_2}) and (\ref{equ:lambdacondition2}).

Fig.~\ref{fig:suscomega} shows the frequency dependence of the momentum-integrated charge and spin susceptibilities obtained by three different methods.
The red squares indicate the \lamdm results where both the charge and the spin susceptibilities have been renormalized by a $\lambda$-correction.
They are compared to the corresponding DMFT results where $\ldens=\lmagn=0$ (orange circles) and the \lamm where only the spin susceptibility is corrected using Eq.~(\ref{equ:lambdacondition1}) (green hexagons). For further comparison, we also present the local impurity susceptibilities $\chi^{\omega}_{r,\text{loc}}$ (brown triangles) which have been obtained directly from the DMFT impurity solver.
We show our data for three different values of $U$ in the weak ($U=1$), intermediate ($U=2$) and strong ($U=3.5$) coupling regimes at temperatures slightly above the DMFT phase transition. 

We observe that the introduction of $\lambda$-corrections leads to an overall suppression of the charge and spin susceptibilities with respect to DMFT in the entire parameter regime. 
For the spin susceptibility in the upper panels, this reduction becomes more pronounced upon increasing $U$. This observation can be attributed to the overall increase of spin fluctuations by the gradual emergence of a local moment for larger interaction values.
In fact, the absence of two-particle self-consistency in DMFT leads to a substantially larger violation of the sum rule Eq.~(\ref{equ:lambdacondition1}) when local spin fluctuations enhance the nonlocal spin susceptibility~\cite{DelRe2021}. 
Let us remark that the renormalization of the spin susceptibility becomes also stronger when the temperature is decreased.
This can be readily understood by the substantial growth of this correlation function upon approaching $\TN$ of DMFT (where it actually diverges) leading to a stronger violation of Eq.~(\ref{equ:lambdacondition1}). 

Let us now address the difference in the spin renormalization between the \lamm and \lamdm methods.
The reduction is larger for \lamm where {\em only} the spin fluctuations are renormalized by means of Eq.~(\ref{equ:lambdacondition1}) (green hexagons) compared to \lamdm where we consider a $\lambda$-correction in both the spin {\em and} the charge channel (red squares).
As discussed in the previous Sec.~\ref{sec:determinationlambda}, this behavior can be understood from Eq.~(\ref{equ:lambdacondition1}) where a suppression of $\cdensF$ through $\ldens>0$ must be compensated by a smaller value of $\lmagn$ and, hence, a larger $\cmagnF$ compared to \lamm where $\ldens\!=\!0$, to match the constant on the right-hand side of this equation. 
This effect is more pronounced at weak coupling ($U=1$, upper left panel) and gradually decreases upon increasing $U$. In fact, while at intermediate coupling ($U=2$, upper middle panel) the difference between \lamm and \lamdm is still visible (albeit very small) both methods provide virtually the same result at strong coupling ($U=3.5$, upper right panel).

On the contrary, the (relative) change of the charge susceptibility due to the introduction of $\ldens$ is rather constant (about 30\%) in the entire parameter regime (see difference between red squares and orange circles in the lower panels of Fig.~\ref{fig:suscomega}). However, the effect of this renormalization on other physical quantities strongly depends on the coupling strength. At weak coupling ($U=1$, lower left panel), charge fluctuations are still significant (compared to the value of the spin fluctuations) and their correction by means of $\ldens$ is indeed highly relevant for the fulfillment of sum rule~(\ref{equ:lambdacondition1}). 
In fact, at $U=1$, the charge renormalization is almost solely responsible for the enforcement of this consistency relation as the spin susceptibility is more or less equivalent to DMFT (cf.~red squares and orange circles in the left upper panel of Fig.~\ref{fig:suscomega}).
Upon increase of the interaction strength to $U=2$ and $U=3.5$, the overall size of the charge susceptibility decreases by one to two-orders of magnitude.
Hence, the effect of the charge correction on the spin renormalization becomes gradually smaller and almost vanishes in the strong coupling regime where only the spin susceptibility contributes significantly to Eq.~(\ref{equ:lambdacondition1}). 

\begin{figure*}[t!]
   \centering
    \includegraphics[width=0.48\linewidth]{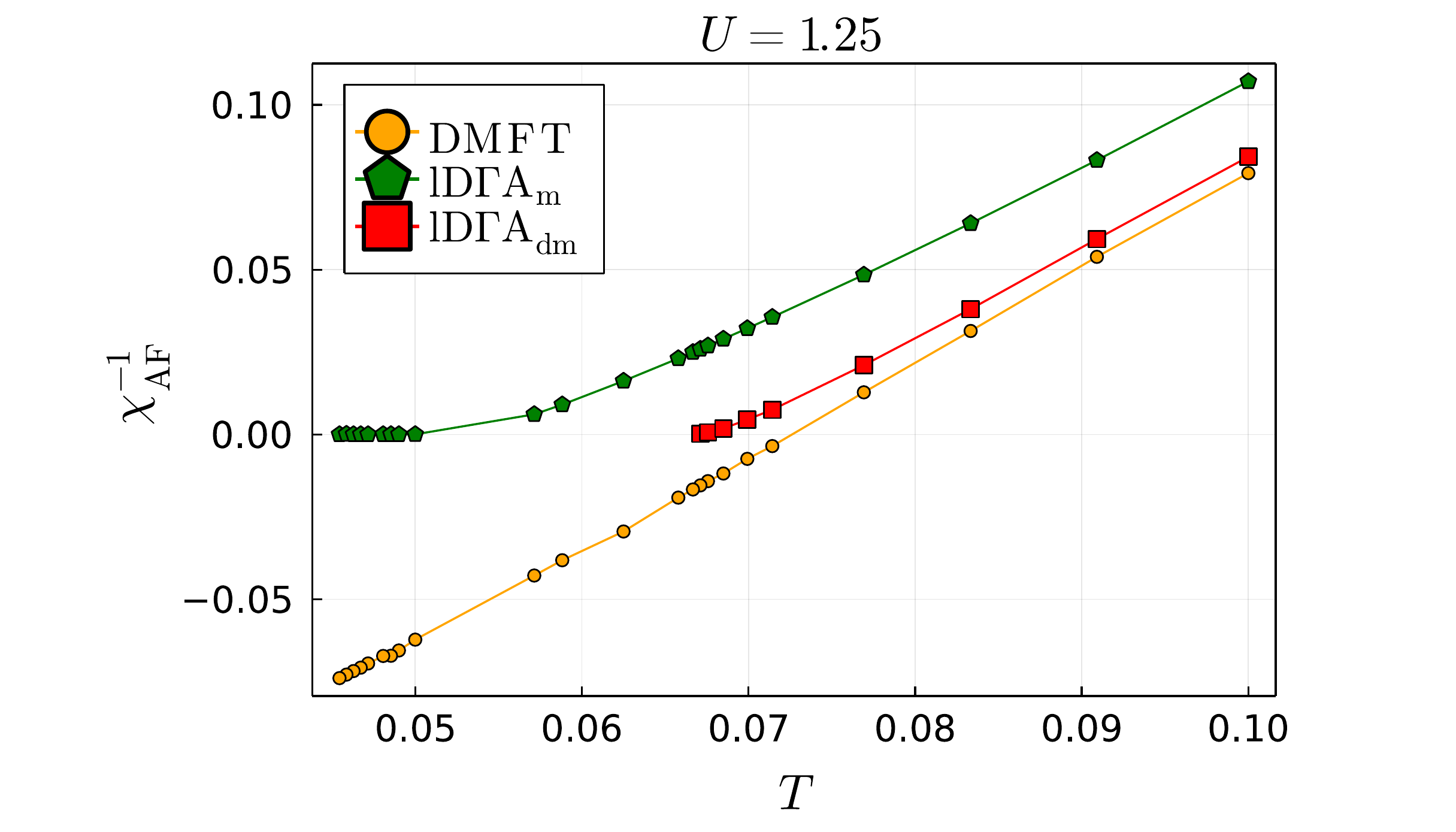}
    \includegraphics[width=0.48\linewidth]{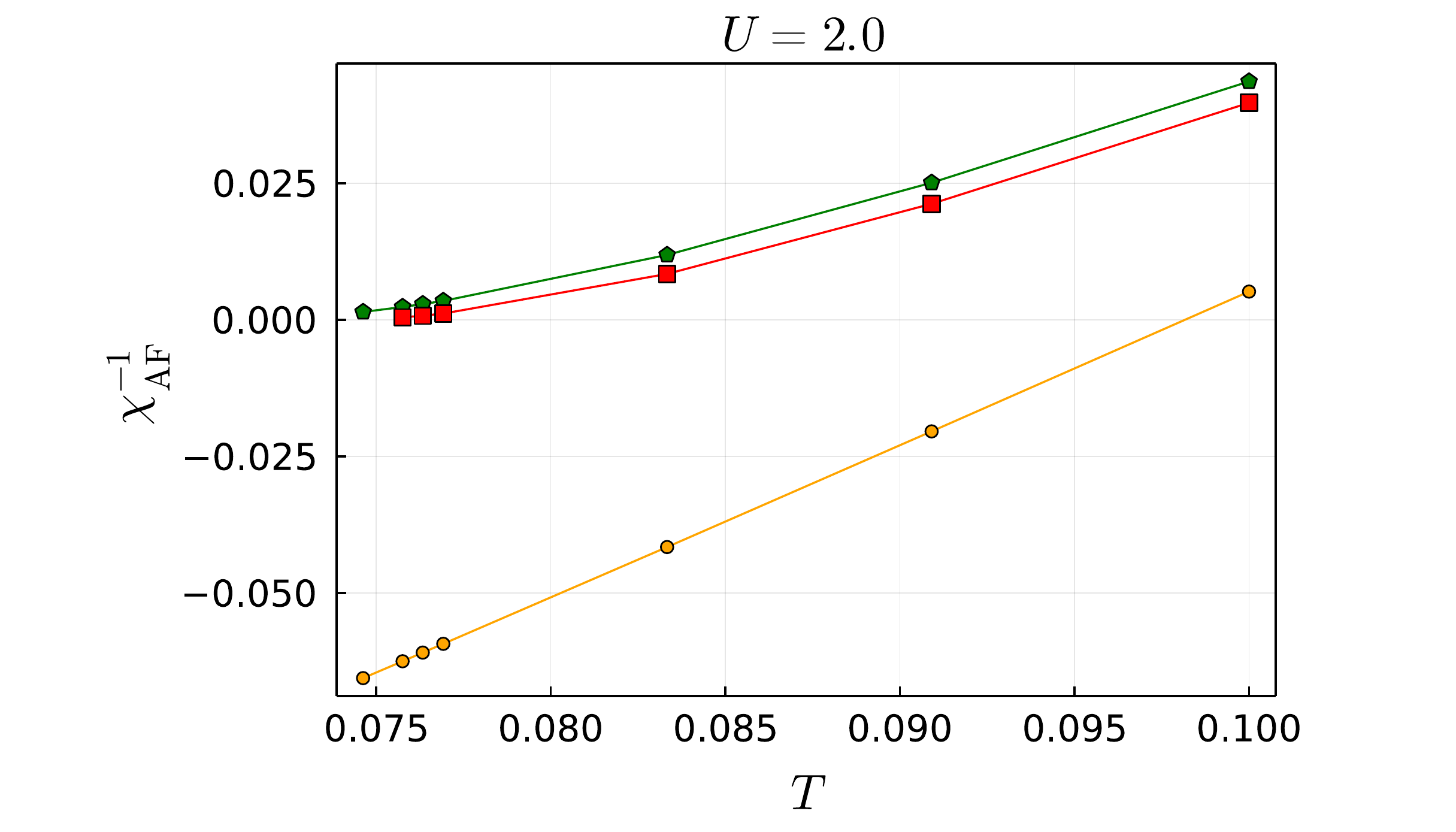}
    \caption{Inverse of the antiferromagnetic susceptibility $\chi_\text{AF}(T)=\cmagn(\omega\!=\!0,\mathbf{q}\!=\!\mathbf{\Pi})$ as a function of the temperature $T\!=\!\frac{1}{\beta}$ for $U\!=\!1.25$ (left) and $U\!=\!2$ (right) obtained by DMFT (orange circles), \lamm (green hexagons)  and \lamdm (red squares).}
    \label{fig:chiAF}
\end{figure*}

It is also instructive to consider the deviations of the momentum-summed lattice susceptibilities of DMFT  (orange circles) from the local ones of the AIM related to the DMFT solution of the Hubbard model (brown triangles in Fig.~\ref{fig:suscomega}).
Since DMFT is not a two-particle self-consistent theory considerable differences between these quantities are to be expected. 
This is indeed true for the spin channel (upper panels), while no significant (relative) difference can be observed in the charge channel. 
Introducing a $\lambda$-correction solely in the spin channel (green hexagons) we observe that the consistency between the momentum summed lattice susceptibility and the local one of the AIM is implicitly restored.
While at a first glance this effect from the \lamm method appears to be preferable, we argue that in fact the opposite is the case. 
The local impurity model of DMFT contains no nonlocal correlation effects. 
Hence, its local correlation functions are expected to deviate from the local part of the corresponding D$\Gamma$A lattice correlation functions which indeed contain such nonlocal contributions.
These nonlocal contributions can be included in an impurity model only by introducing an effective frequency dependent interaction $U(\omega)$ as done in the dual boson approach~\cite{Rubtsov2012}.
In this method, a consistency between local lattice quantities and the corresponding impurity quantities is indeed meaningful because nonlocal correlation effects are partially encoded in the frequency dependence of the effective $U$. 
Since we do not consider such a modification of the impurity model within the ladder D$\Gamma$A, a consistency between momentum summed and impurity correlation function at the two-particle level is not to be expected.
An additional $\lambda$-correction in the charge channel leads to physically reasonable deviations from the correlation function of the AIM (red squares versus brown triangles in Fig.~\ref{fig:suscomega}).

\subsection{Phase diagram}\label{sec:phasediagram}

\begin{figure}[t!]
    \centering
    \includegraphics[width=\linewidth]{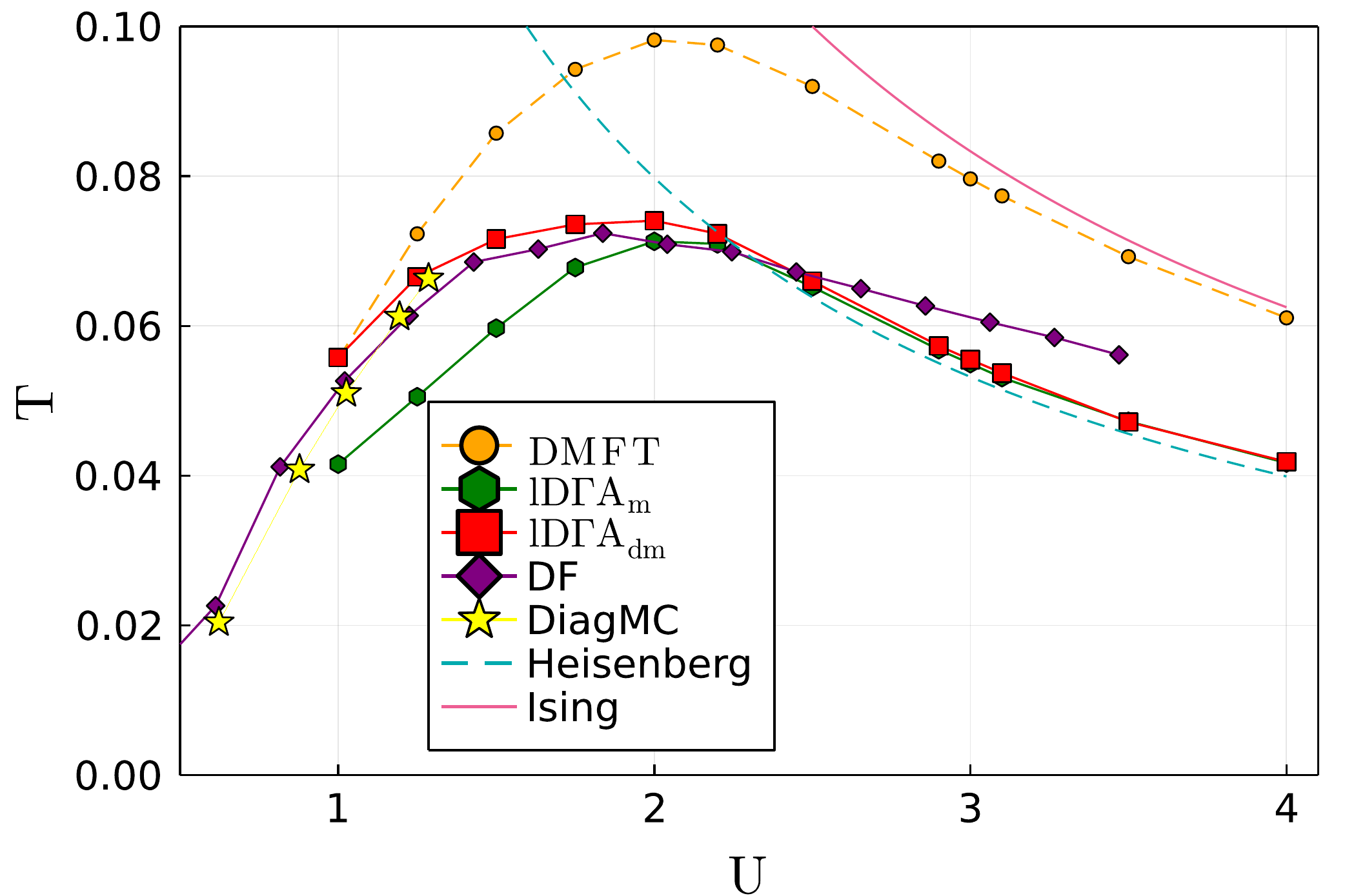}
    \caption{Phase diagram of the $3d$ half-filled Hubbard model on a simple cubic lattice with nearest neighbor hopping. The curves correspond to the transition temperature to the antiferromagnetically ordered state obtained by the different methods (DF~\cite{Hirschmeier2015} and DiagMC~\cite{PhysRevLett.129.107202}) indicated in the legend of the figure.}
    \label{fig:pd_res}
\end{figure}

In three dimensions, the half-filled Hubbard model on a bipartite simple cubic lattice features an antiferromagnetically ordered phase at low temperatures for all values of the interaction parameter $U$.
The second-order phase transition to this antiferromagnetic state is indicated by the divergence of the antiferromagnetic susceptibility $\chi_\text{AF}(T)\!=\!\cmagn(\omega\!=\!0,\mathbf{q}\!=\!\mathbf{\Pi})$. 
In Fig.~\ref{fig:chiAF}, we  present the results for the {\em inverse} of this observable as a function of the temperature for two different values of $U$. 
A divergence of $\chi_\text{AF}(T)$, i.e., vanishing of $\chi_\text{AF}^{-1}(T)$ marks the onset of antiferromagnetic order.
We observe the same hierarchy of curves as in the previous section.
The DMFT antiferromagnetic susceptibility (orange circles), which corresponds to $\ldens=\lmagn=0$, is larger than the D$\Gamma$A susceptibilities (green hexagons and red squares) where $\lambda$-corrections have been applied. 
Consistent with the discussion above, the \lamm results where $\ldens=0$ (green hexagons) are smaller than the ones obtained by \lamdm where both the charge and the spin channels are renormalized (red squares). 
As has been detailed in Sec.~\ref{sec:susceptibilities}, this is explained by the consistency relation~(\ref{equ:lambdacondition1}) where the suppression of charge fluctuations by $\ldens>0$ requires a larger spin susceptibility compared to the case where $\ldens=0$.
The difference between the two approaches is particularly pronounced at weak and intermediate coupling while it gradually decreases for increasing $U$ when charge fluctuations are strongly suppressed and, hence, have lesser effect on the overall physical picture.

Close to the transition temperature $\TN$, $\chi_\text{AF}(T)$ takes the form of a universal scaling function~\cite{Binney1992}
\begin{equation}\label{equ:scalingfunction}
    \chi_\text{AF}(T)\sim a\lvert T-\TN\rvert^{-\gamma},
\end{equation}
where $\gamma$ is the critical exponent associated with the susceptibility. 
The mean-field (MF) value $\gamma=\gamma_\text{MF}=1$ is consistent with the linear temperature dependence of the DMFT $\chi_\text{AF}^{-1}(T)$ in Fig.~\ref{fig:chiAF} (orange circles). 
The renormalization of this DMFT susceptibility by a $\lambda$ parameter leads to a modification of the mean-field behavior and provides a $\gamma>1$ which is clearly visible for the green hexagons and red squares in Fig.~\ref{fig:chiAF}. 
The deviation from the linear mean-field behavior can be only observed in the critical temperature region $\Delta T_\text{crit}\propto \TN^2$ according to the Ginzburg criterion~\cite{Landau1980}.
This explains why the bending of $\chi_\text{AF}^{-1}(T)$ is more pronounced in a wider temperature range for $U=2$, where $\TN$ is substantially larger than for $U=1.25$ (for lower values of $U$ the critical regime is hardly visible on our scales). 
In Ref.~\onlinecite{DelRe2019}, it has been discussed, that the \lamm provides critical exponents consistent with the spherical symmetric Kac model~\cite{Stanley1971}, where $\gamma=2$, similar as in TPSC~\cite{Dare1996}. 
On the other hand, when the susceptibility bends away from the mean-field behavior, $\gamma\approx 1.4$ as in the Heisenberg model has been fitted numerically in an extended temperature range~\cite{Rohringer2011}. Such an exponent has also been observed in the DF approach~\cite{Hirschmeier2015}, whereas that for the Falicov Kimball model was consistent with the critical exponent of the Ising model~\cite{Antipov2014}.
Let us point out that fitting the exponent of a scaling function such as Eq.~(\ref{equ:scalingfunction}) is intrinsically difficult and $\gamma=2$ can be only achieved by including subleading terms in the fit as has been shown in Refs.~\onlinecite{Semon2012,DelRe2019}.

In any case, the determination of $\TN$ from numerical data is stable, and its value depends only very weakly on changes in $\gamma$~\cite{DelRe2019}.
We have, hence, fitted the results for $\chi_\text{AF}(T)$ to the scaling function in Eq.~(\ref{equ:scalingfunction}) in order to obtain the transition temperature $\TN$ for different interaction values $U$.
The transition curves $\TN(U)$ for DMFT (orange circles), the \lamm (green hexagons), and \lamdm (red squares) are depicted in Fig.~\ref{fig:pd_res} where also results obtained with other methods are shown for comparison. 
Overall, a reduction of $\TN$ obtained by both versions of D$\Gamma$A with respect to the DMFT curve can be observed.
This is indeed the expected behavior as mean-field theories (such as DMFT) typically overestimate the transition temperature to an ordered state.
This can be attributed to the fact that nonlocal correlations, which are included in D$\Gamma$A in an effective way by the $\lambda$-corrections but not in DMFT, destroy the order in an intermediate temperature regime and predict a reduced $\TN$.
Remarkably, in the weak to intermediate coupling region ($U\!\sim\!1$ to $U\!\sim\!2$) this reduction is much more pronounced when only the renormalization of the spin susceptibility through Eq.~(\ref{equ:lambdacondition1}) is taken into account (green hexagons).
This is a direct consequence of the mechanism which has been discussed in Sec.~\ref{sec:susceptibilities} for the susceptibilities:
The positive $\ldens$ leads to a decrease of the charge susceptibility $\cdensF$ in Eq.~(\ref{equ:lambdacondition1}) requiring a larger spin susceptibility $\cmagnF$ (corresponding to a smaller value of $\lmagn$ with respect to the case where only the spin channel is corrected).
Consequently, the related antiferromagnetic spin susceptibility $\chi_\text{AF}(T)$ will diverge at a higher temperature $T$ in \lamdm giving rise to a higher transition temperature $\TN$ with respect to \lamm.

For $U\lesssim 1$, our numerical \lamdm data for $\TN$ coincide with the corresponding DMFT results.
This means nonlocal correlations do not reduce the transition temperature in this parameter regime,
which is indeed the expected behavior and has been predicted by analytical considerations and numerical simulations~\cite{Tahvildar-Zadeh1997,Schauerte2002,DelRe2021}. In fact, it was demonstrated that $\TN$ is affected mainly by local particle-particle fluctuations (which are of course already included in DMFT) in the weak coupling region. 
In the intermediate coupling regime ($U\!\sim\!1$ to $U\!\sim\!2$) we observe a reduction of $\TN$ in \lamdm with respect to DMFT which is in good agreement with dual fermion (DF)~\cite{Hirschmeier2015} and diagrammatic Monte Carlo~\cite{PhysRevLett.129.107202} results.
This is consistent with the fact that within the DF treatment of the problem both the spin and the charge fluctuations are renormalized within a self-consistent update of the generalized susceptibilities in the dual space~\cite{Rohringer2018a} (although a consistency of the potential energy has not been demonstrated in this framework).
Diagrammatic Monte Carlo calculations provide (in principle) the exact solution of the problem. 
In the intermediate coupling region, they are in very good agreement with our \lamdm results which can therefore be considered a more reliable method than \lamm for the estimation of the transition temperature in this parameter regime.% {\color[rgb]{1,0,0}GR:Check with the new DiagMC data!}

%they predict a $\TN$ between the ones of \lamdm and \lamm indicating that our new approach might slightly overestimate $\TN$ in this coupling regime.
%in good agreement with \lamdm which can therefore be considered a more reliable method than \lamm for the estimation of the transition temperature in this parameter regime. 
% Doppelt? as which underestimates $\TN$ due to a lack of charge renormalization.

Finally, in the strong coupling region $U\!\gtrsim\!2$, the results of both D$\Gamma$A schemes coincide and match excellently the data from the Heisenberg model onto which the Hubbard model can be mapped at large interaction strength.
This is consistent with the fact, that in this parameter region the charge degrees of freedom are almost frozen and, hence, the renormalization in the charge channel has no effect on $\TN$.

\subsection{Self-energies}\label{sec:resultssigma}

\begin{figure*}[t!]
   \centering
     \subfloat[\label{fig:sigma_125}]{
         \centering
         \includegraphics[width=0.31\linewidth]{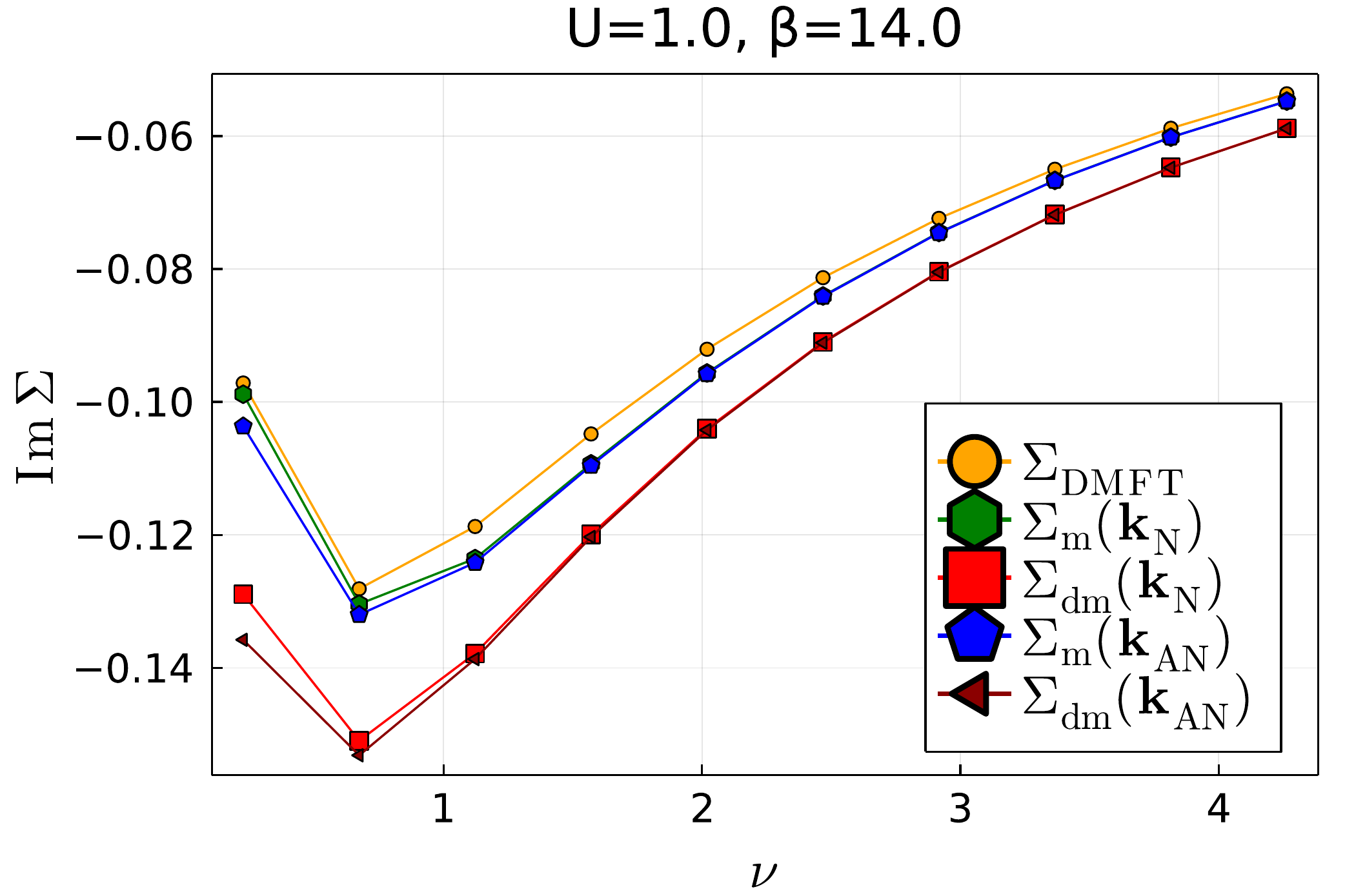}
     }
     \hfill
     \subfloat[\label{fig:sigma_20}]{
         \centering
         \includegraphics[width=0.31\linewidth]{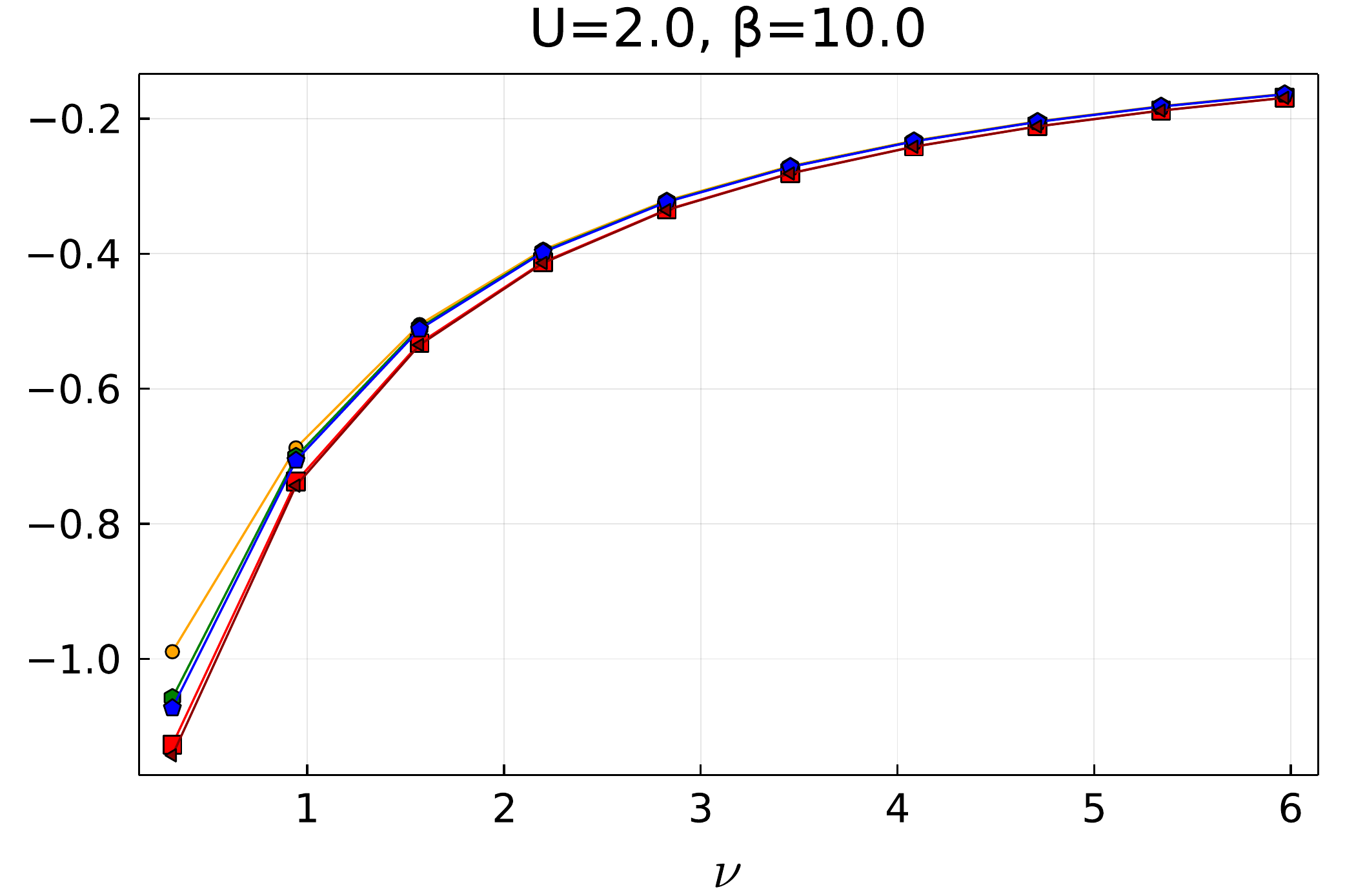}
     }
     \hfill
     \subfloat[\label{fig:sigma_u38}]{
         \centering
         \includegraphics[width=0.31\linewidth]{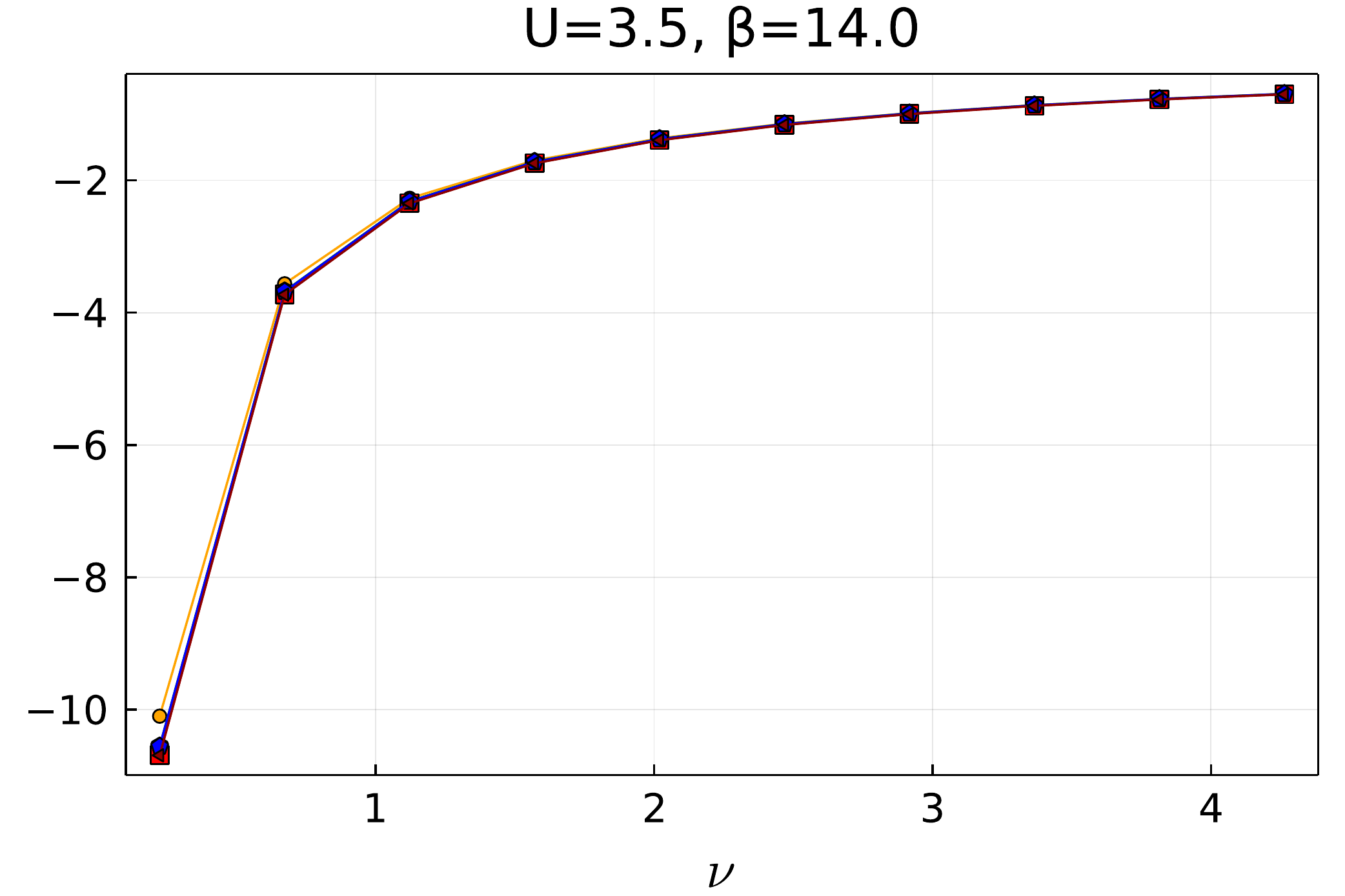}
    }
     \hfill
     \caption{Imaginary part of the electronic self-energy at the nodal [$\bk_\text{N}=(\frac{\pi}{2},\frac{\pi}{2},\frac{\pi}{2})$] and antinodal [$\bk_\text{AN}=(\pi,\frac{\pi}{2},0)$] points on the Fermi surface for three different coupling strengths at $\beta=10$ and $\beta=14$ as a function of the fermionic Matsubara frequency $\nu$. We present data for DMFT (orange circles), \lamm (green hexagons and blue pentagons) and \lamdm (red squares and brown triangles).}
    \label{fig:sigma_comparison}
\end{figure*}

\begin{figure}
         \centering
         \includegraphics[width=\linewidth]{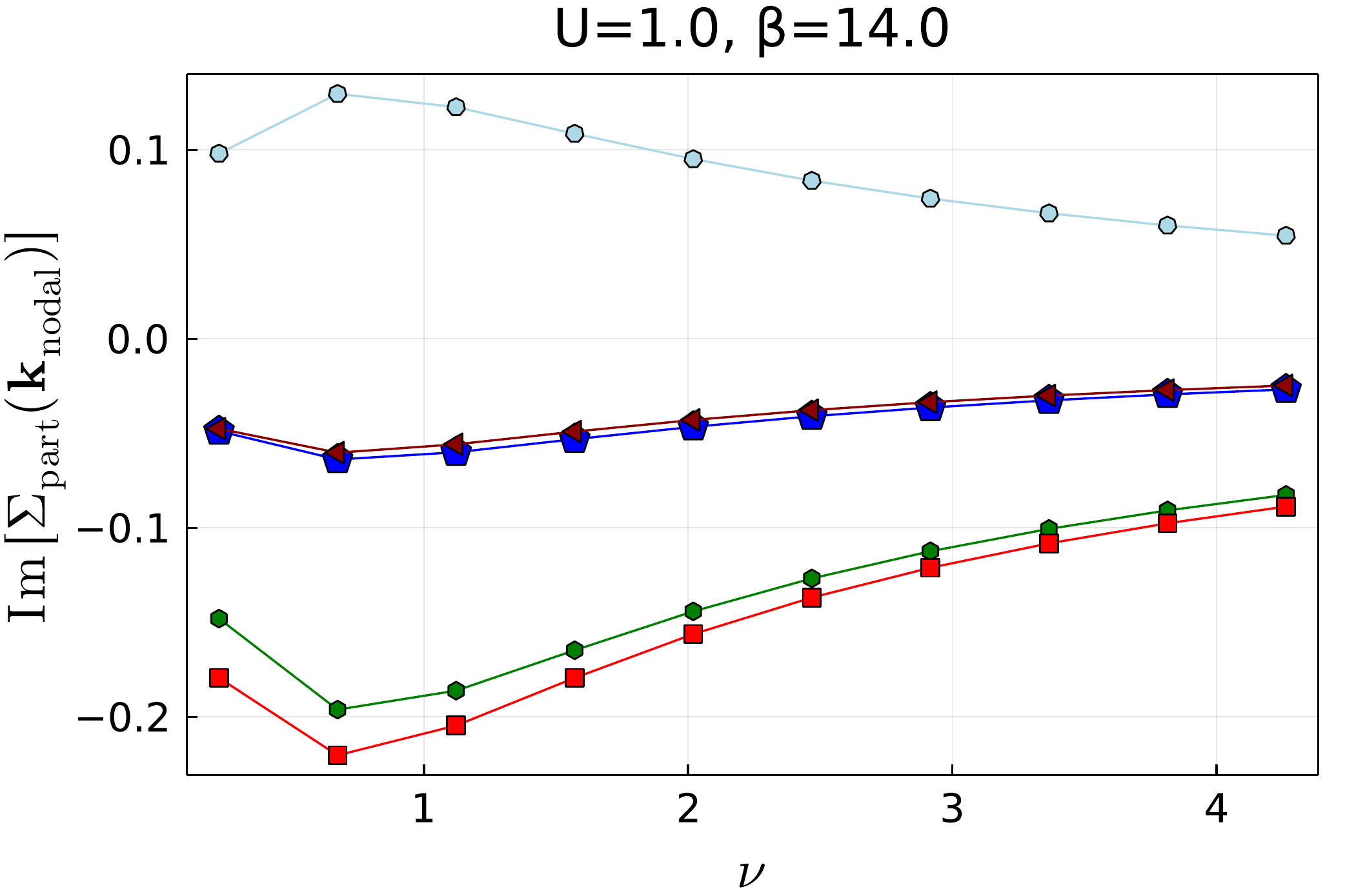}
     \caption{Imaginary part of the self-energy $\Sigma^{\nu}_{\bk_\text{N}}$ split into contributions from the charge susceptibility (blue pentagons and brown triangles), the spin susceptibility (green hexagons and red squares) and an remainder (blue heptagons) according to Eqs.~(\ref{subequ:splitsigma}) at $\beta=14$ and $U=1$. Results are presented for \lamdm (red squares and brown triangles) and \lamm (green hexagons and blue pentagons). Note that the remainder is equivalent for both methods as it does not depend on any $\lambda$ parameter.}
    \label{fig:sigma_channelsweak}
\end{figure}
\begin{figure}[t!]
         \centering
         \includegraphics[width=\linewidth]{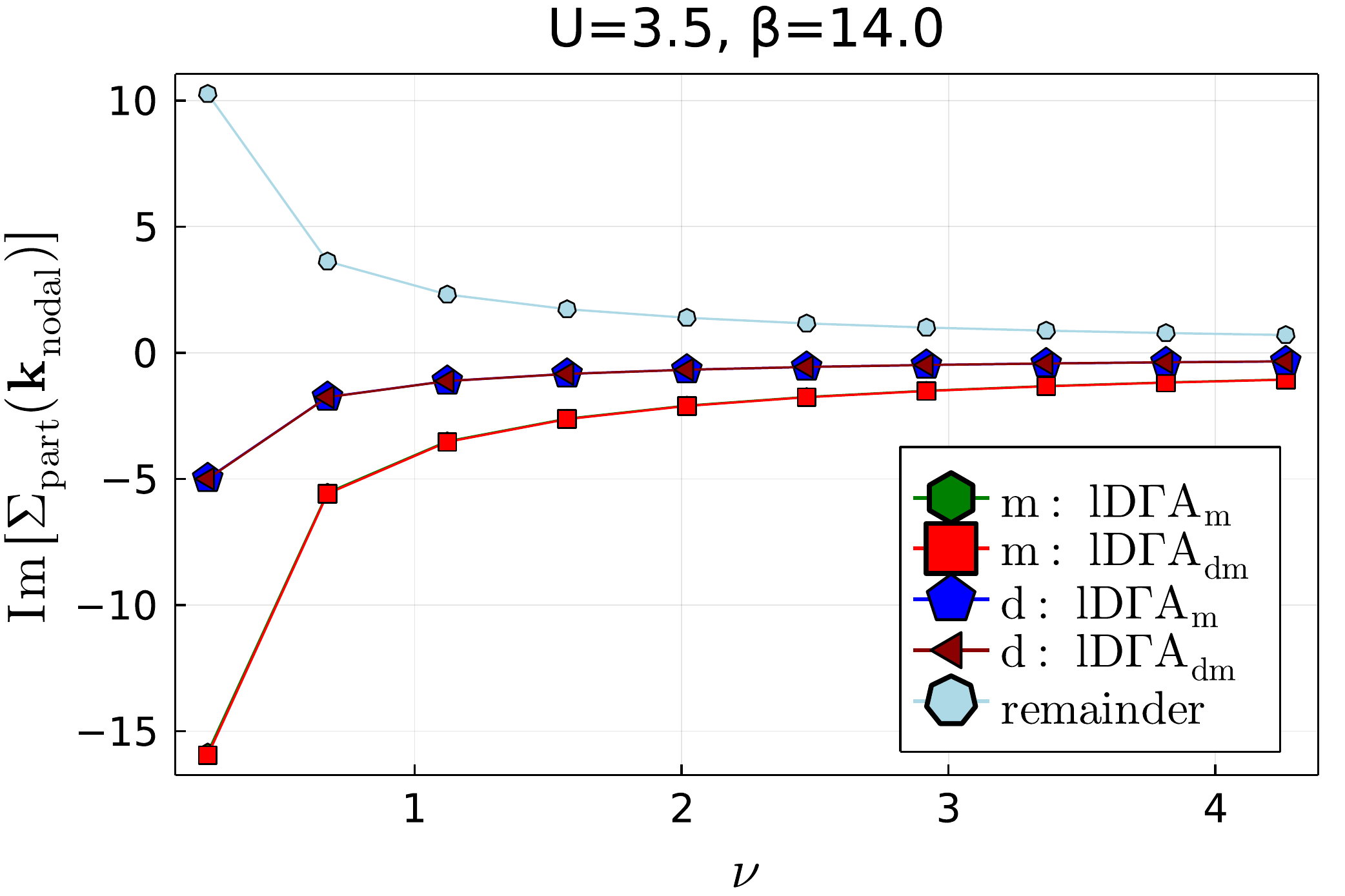}
     \caption{Same as in Fig.~\ref{fig:sigma_channelsweak} but for $U=3.5$.}
    \label{fig:sigma_channelsstrong}
\end{figure}
%
%Imaginary part of self-energy for $U=3.5$, split into contributions. See also Fig.~\ref{fig:sigma_channelsweak}.
In this section, we discuss the momentum dependent imaginary part of the electronic self-energies obtained by \lamdm as a function of the Matsubara frequency $\nu$ at the nodal $\bk_\text{N} = (\pi/2,\pi/2,\pi/2)$ and antinodal $\bk_\text{AN} = (\pi,\pi/2,0)$ momentum on the Fermi surface.
In Fig.~\ref{fig:sigma_comparison}, we compare our findings (red squares and brown triangles) to the corresponding local self-energy $\Sigma^\nu$ of DMFT (orange circles) and and to \lamm results (green hexagons and blue pentagons) for three different values of $U$ at $\beta=10$ and $\beta=14$, slightly above $\TN$ of DMFT for the respective $U$ values.
In general, the absolute values of the momentum dependent self-energies in both D$\Gamma$A schemes are larger than the corresponding DMFT correlation function.
This is the expected behavior~\cite{Rohringer2016} as nonlocal correlations typically suppress the spectral weight at the Fermi level (in addition to the suppression due to local correlations which are already captured by DMFT). 
As has been discussed in several previous papers~\cite{Rohringer2016,Rohringer2018a}, the enhancement of $\Sigma^\nu_\bk$ is stronger at the antinodal point $\bk_\text{AN}$ than at the nodal point $\bk_\text{N}$ which is confirmed by our data (for both variants of D$\Gamma$A). 
We observe that the \lamdm method yields larger (in absolute value) self-energies than the ones obtained by the \lamm method over the entire $U$~range.
This can easily be understood from the different magnitudes of the spin and charge susceptibilities in both approaches and the way how they enter in the EoM~(\ref{equ:eom_2}). 
To this end, we split Eq.~(\ref{equ:eom_2}) into a magnetic contribution, a density contribution and a remainder which accounts for the remaining terms on the right-hand side of this equation:
\begin{subequations}
\label{subequ:splitsigma}
\begin{align}
    \sdensF & =\frac{U^2}{2}\sum_{\omega\mathbf{q}}\gamma^{\nu\omega}_{d,\mathbf{q}}\chi_{\text{d},\mathbf{q}}^{\lambda_\text{d},\omega} G^{\nu+\omega}_{\bk+\bq}, \label{equ:splitsigmacharge} \\
\smagnF & =\frac{3U^2}{2}\sum_{\omega\bq}\gamma^{\nu\omega}_{\text{m},\bq} \chi_{\text{m},\mathbf{q}}^{\lambda_\text{m},\omega} G^{\nu+\omega}_{\bk+\bq}, \label{equ:splitsigmaspin} \\
\Sigma^{\nu}_{\text{rem},\bk} & =\Sigma^\nu_{\bk}-\Sigma^\nu_{\mathrm{d}, \bk}-\Sigma^\nu_{\mathrm{m}, \bk}
\end{align}
\end{subequations}
Note that the remainder term does not contain $\chi_{\text{d},\mathbf{q}}^{\lambda_\text{d},\omega}$ or $\chi_{\text{m},\mathbf{q}}^{\lambda_\text{m},\omega}$ and therefore does not depend on any $\lambda$ parameter.
It is for this reason equivalent in the \lamm and \lamdm method.
As discussed in Sec.~\ref{sec:susceptibilities} $\cmagnF$ is larger for \lamdm than for \lamm.
This property is directly transferred to $\smagnF$ in Eq.~(\ref{equ:splitsigmaspin}) where at weak coupling ($U=1$) we indeed observe a larger contribution of spin fluctuations to the self-energy for \lamdm (red squares) compared to \lamm as shown in Fig.~\ref{fig:sigma_channelsweak}. 
The opposite behavior is observed for the charge susceptibility. 
It is smaller for \lamdm with respect to \lamm and the same behavior is observed for the corresponding contribution to the self-energy $\sdensF$.
However, since the charge fluctuations are substantially smaller than the spin fluctuations, the former are less relevant in the equations of motion which leads to an overall larger self-energy for $\ldens>0$.
The same should in principle hold in the strong coupling regime.
However, as discussed in the previous sections, due to the extremely small values of charge fluctuation also the differences in the self-energies and their various contributions are strongly suppressed and almost no differences in the results for \lamdm and \lamm can be observed as it is shown in Fig.~\ref{fig:sigma_channelsstrong}.

Let us remark that at weak-to-intermediate coupling the difference for $\Sigma_\mathbf{k}^\nu$ between our new \lamdm method and the previous \lamm approach is quite substantial. As we observe a tiny deviation from the correct high-frequency asymptotic behavior $\Sigma_\mathbf{k}^\nu\underset{\nu\rightarrow\infty}{=}U^2\frac{n}{2}\left(1-\frac{n}{2}\right)\frac{1}{i\nu} + \mathcal{O}(\frac{1}{i\nu^2})$ for \lamdm, our new approach might slightly overestimate this correlation function. However, at present, a comprehensive understanding of this feature is still lacking and further investigation in future research work is required.

\subsection{Potential and kinetic energies}\label{sec:epot}

\begin{figure*}[t!]
    \centering
    \includegraphics[width=0.46\linewidth]{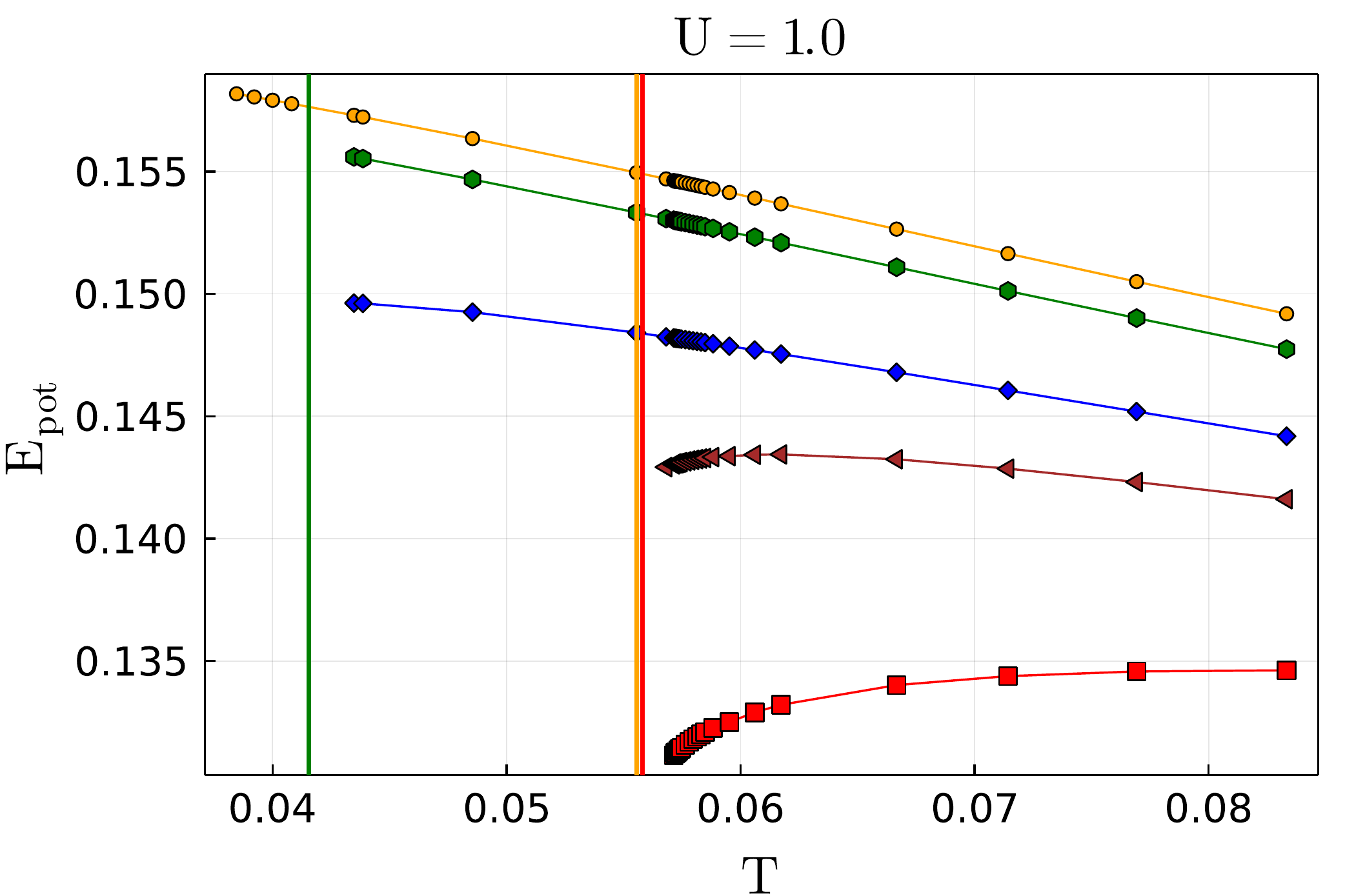}
    \includegraphics[width=0.46\linewidth]{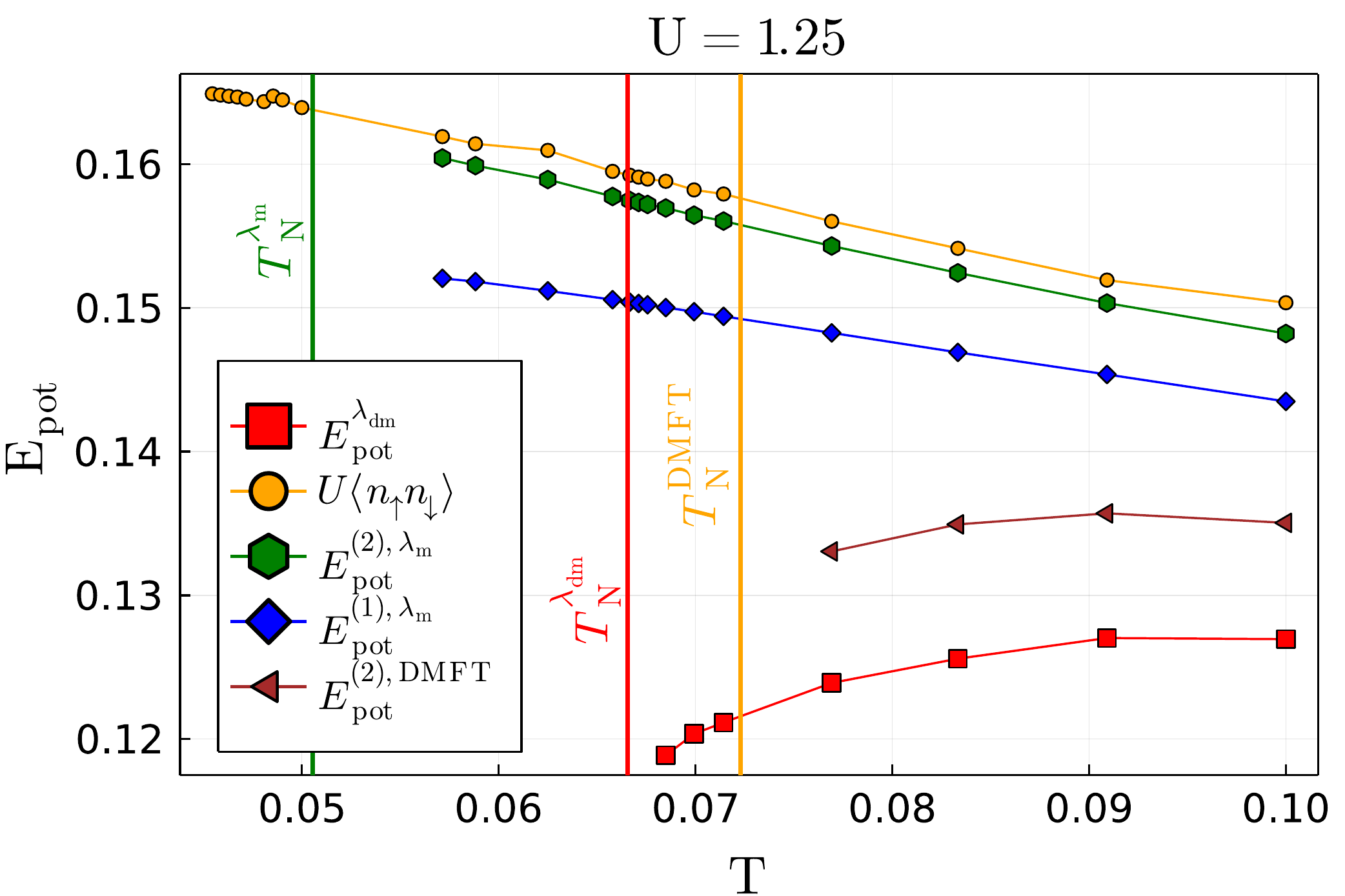}\hfill
    \includegraphics[width=0.46\linewidth]{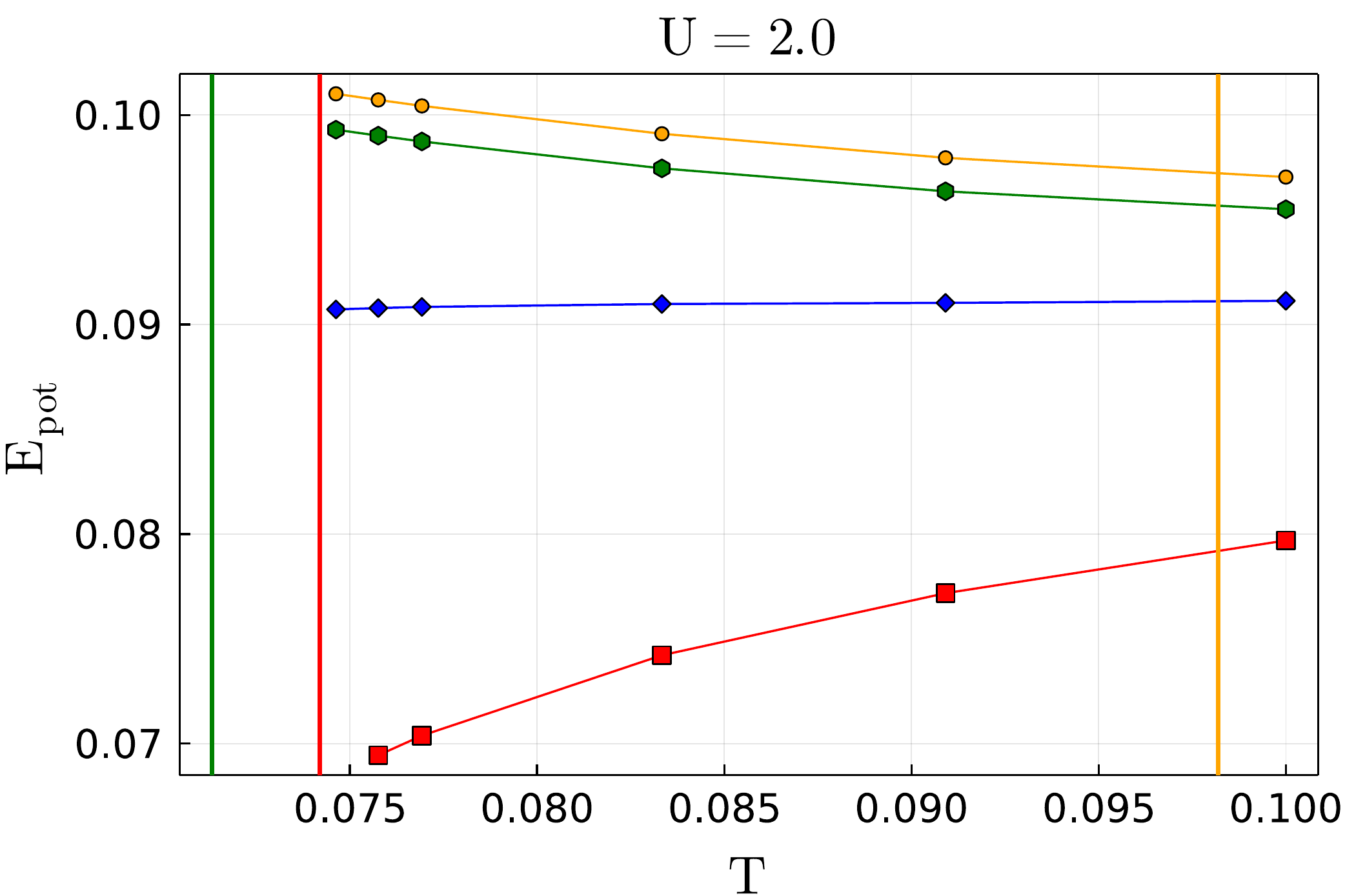}
    \includegraphics[width=0.46\linewidth]{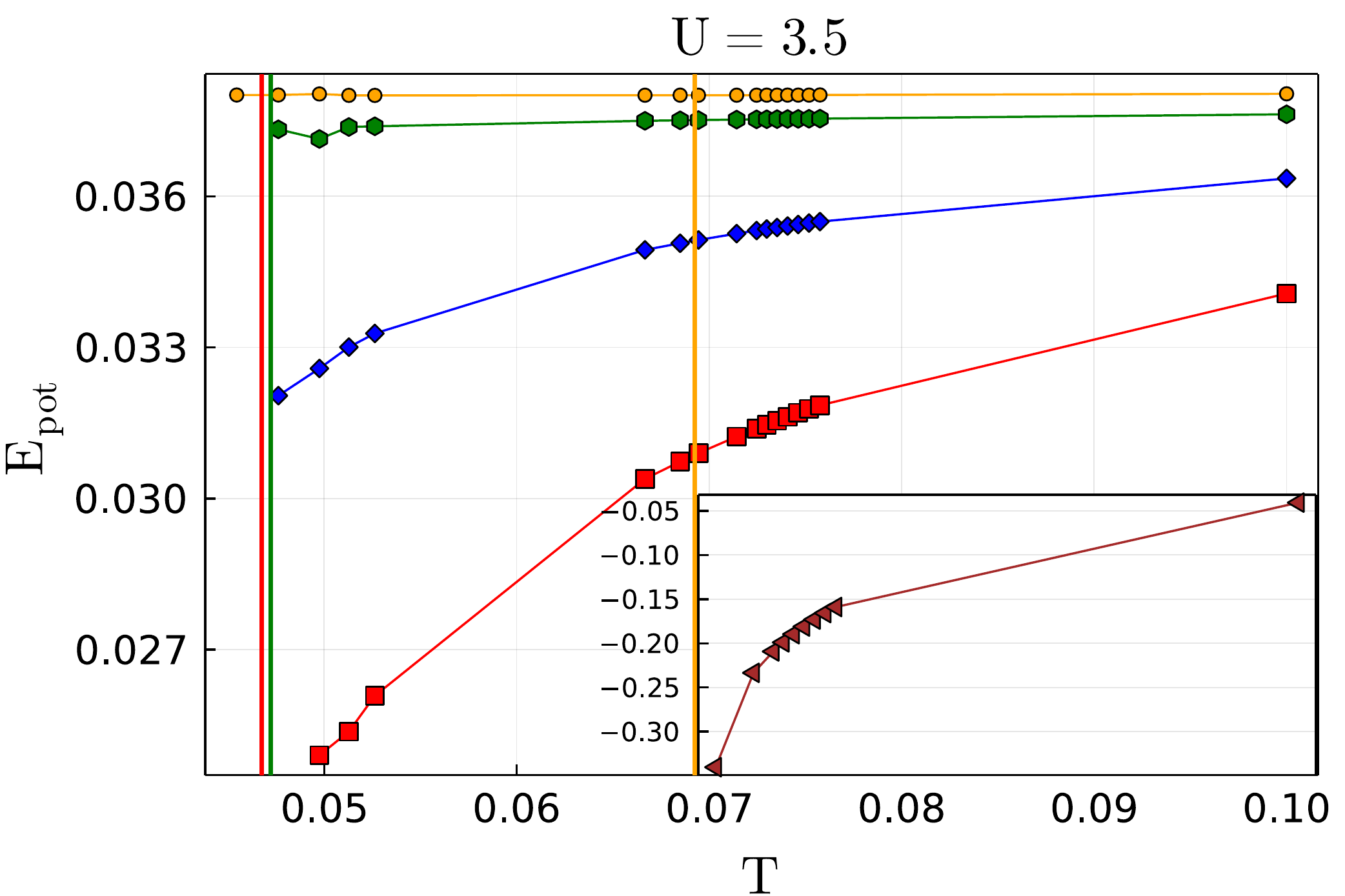}
    \caption{Potential energy as a function of temperature for four different values of $U$ as obtained by the DMFT (orange circles and brown triangles), \lamm (blue diamonds and green hexagons), and \lamdm (red squares). Note that for DMFT and \lamm the results for $E_\text{pot}$ calculated from the one- and the two-particle levels [corresponding to the right- and left-hand sides of Eq.~(\ref{equ:lambdacondition2}), respectively] differ. Vertical lines indicate the transition temperature $\TN$ of the respective method. Due to the unphysical large scale of $E_\text{pot}^{(2)}$ of DMFT at $U=3.5$, the data are shown as inset. For $U=2$ (lower left panel) $E_\text{pot}^{(2)}$ of DMFT (brown triangles) is not shown, because it can only be calculated above $\TN$ of DMFT (orange line), i.e., only for $T=0.1$ in this figure. The corresponding value at this temperature is $E_\text{pot}^{(2)}=-0.0212$.}
    \label{fig:epot}
\end{figure*}
\begin{figure}[htb]
    \centering
    \includegraphics[width=\linewidth]{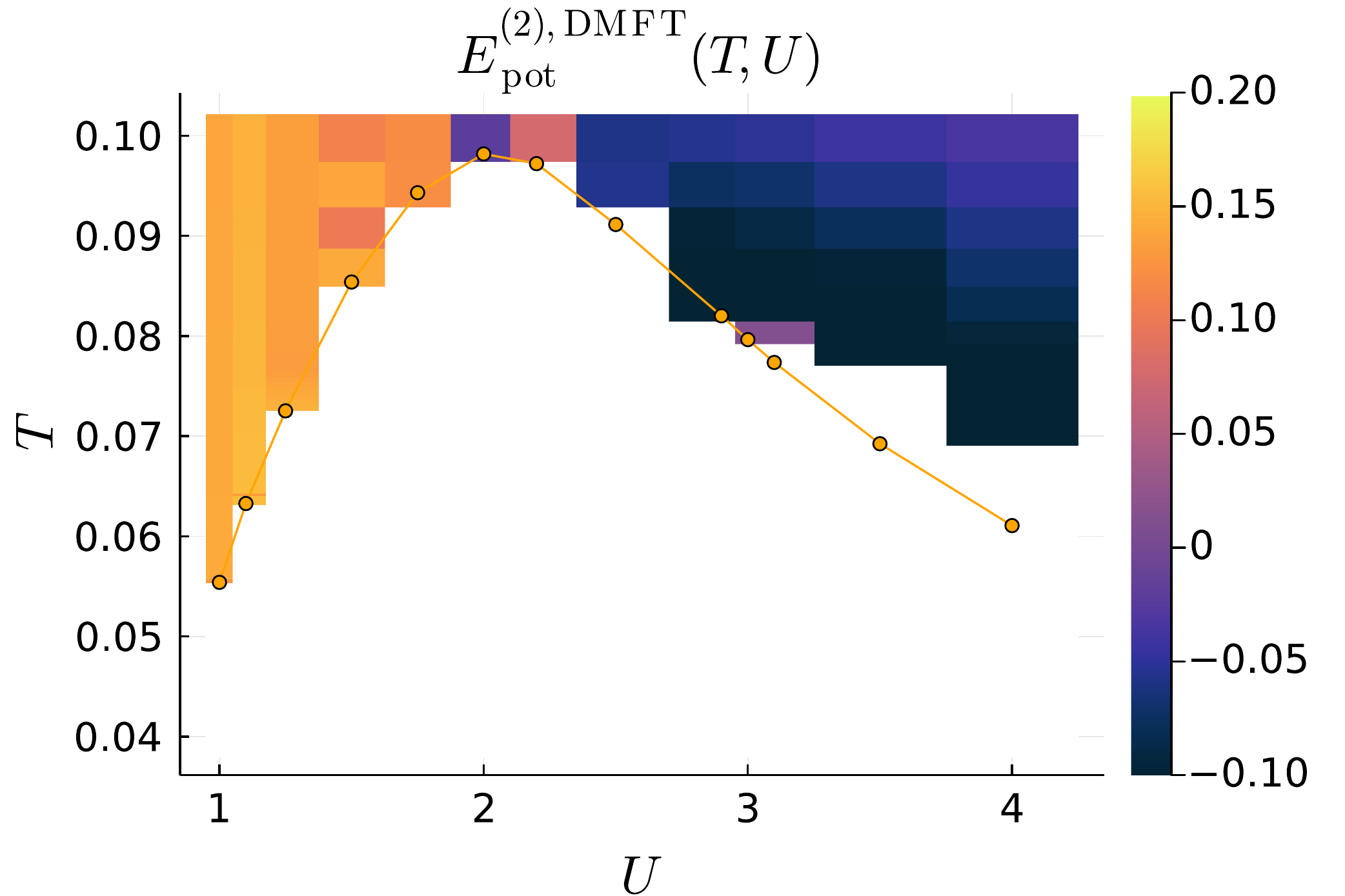}
    \caption{DMFT two particle potential energy $E_\text{pot}^{(2)}$ as a function of $T$ and $U$.}
    \label{fig:Epot_DMFT_2_grid}
\end{figure}
%Note the otherwise rather small range of $E_\text{pot}$.

Figure~\ref{fig:epot} shows the potential energies obtained from DMFT (orange circles and brown triangles), \lamm (blue diamonds and green hexagons), and \lamdm (red squares). 
Let us stress that for the first two cases (DMFT and \lamm) the potential energies obtained at the one particle level ($E_\text{pot}^{(1)}$, orange circles and blue diamonds) deviate from the corresponding results at the two-particle level ($E_\text{pot}^{(2)}$, brown triangles and green hexagons) as these approaches are not two-particle self-consistent [c.f., Eq.~(\ref{equ:lambdacondition2})].
The vertical lines indicate $\TN$ for the respective methods.

We observe the same hierarchy of curves for all values of $U$. The largest potential energy is $E_\text{pot}^{(1)}$ of DMFT (orange circles) which is obtained via the right-hand side of Eq.~(\ref{equ:lambdacondition2}) with the local DMFT self-energy and the (lattice) Green's function of DMFT. 
Considering the DMFT self-consistency relation, we obtain
\begin{align}
\label{equ:epotaim}
    E_\text{pot,DMFT}^{(1)} 
        &= 
        \sum_{\nu} 
            \underset{G^\nu}{\underbrace{\sum_\bk G^\nu_\bk}}
            \Sigma^\nu
        =
        \sum_\nu G^\nu\Sigma^\nu\nonumber\\
        &=
        \frac{U}{2}
        \sum_\omega \left(
            \chi^\omega_\mathrm{d}-\chi^\omega_\mathrm{m} \right)+U\frac{n^2}{4}
        =U\langle n_\uparrow n_\downarrow\rangle,
\end{align}
where $G^\nu_\bk$ is the DMFT lattice Green's function and $G^\nu$ denotes the local impurity Green's function which is equivalent to the local (i.e., momentum summed) DMFT lattice Green's function due to the DMFT self-consistency condition. 
$\chi^\omega_{\text{d}/\text{m}}$ are the local impurity charge and spin susceptibilities and $\langle n_\uparrow n_\downarrow\rangle$ corresponds to the impurity double occupancy. 
Note that the AIM is solved exactly and, hence, all consistency relations, in particular the local version of Eq.~(\ref{equ:lambdacondition2}), are fulfilled as also indicated in Eq.~(\ref{equ:epotaim}).
This implies that $E_\text{pot}^{(1)}$ of DMFT is equivalent to $U$ times the double occupancy of the auxiliary AIM which is obtained directly from the impurity solver.
Alternatively, the same results can be calculated by summing the local susceptibility $\chi^\omega_{\uparrow\downarrow}=\frac{1}{2}[\chi^\omega_\mathrm{d}-\chi^\omega_\mathrm{m}]$ over all bosonic Matsubara frequencies $\omega$.
In Sec.~\ref{sec:susceptibilities}, we have demonstrated that the momentum summed spin and charge lattice susceptibilities in \lamm (green hexagons in the upper and orange circles in the lower panels of Fig.~\ref{fig:suscomega})~\footnote{Note that the momentum summed lattice susceptibility of \lamm in the charge channel is equivalent to the corresponding DMFT result, because no $\lambda$-correction is applied to charge fluctuations (i.e., $\ldens=0$)  within \lamm.} are almost equivalent to the corresponding correlation function of the AIM (brown triangles in Fig.~\ref{fig:suscomega}).
This explains why $E_\text{pot}^{(2)}$ of \lamm (green hexagons) is almost the same (or only very slightly smaller) than the DMFT potential energy $E_\text{pot}^{(1)}$ (orange circles) in Fig.~\ref{fig:epot}.

The D$\Gamma$A values for $E_\text{pot}^{(1)}$ (blue diamonds) on the other hand, are considerably smaller than the corresponding DMFT results which have been also reported in Ref.~\onlinecite{Rohringer2016}. 
In the latter reference it has been discussed, that this is indeed the expected behavior at weak coupling where the antiferromagnetic ground state is of Slater type. 
Within such a weak coupling Slater mechanism, the antiferromagnetic phase is stabilized by a decrease in the potential energy. 
Our results indicate that this mechanism is reflected in the corresponding antiferromagnetic fluctuations above $\TN$ where the inclusion of nonlocal correlations leads to a suppression of $E_\text{pot}$ with respect to DMFT.
We observe the same behavior at strong coupling where, in principle, a reversed order of the hierarchy in magnitude of potential energies could have been expected due to the antiferromagnetic phase being of Heisenberg type and a stabilization through a gain of kinetic energy~\cite{Rohringer2016}.
This is, however, not observed as D$\Gamma$A {\em always} leads to a reduction of the potential energy with respect to DMFT, indicating that this change in the nature of the antiferromagnetic order from weak to strong coupling is not fully captured by the corresponding fluctuations above $\TN$ (see also the discussion of the kinetic energy below). 
Whether this behavior is an artifact of the ladder D$\Gamma$A method or the correct result requires further investigation.

The \lamdm approach (red squares), where by construction $E_\text{pot}^{(1)}=E_\text{pot}^{(2)}$, predicts much smaller potential energies than $E_\text{pot}^{(1)}$ and $E_\text{pot}^{(2)}$ of \lamm (blue diamonds and green hexagons) and $E_\text{pot}^{(1)}$ of DMFT (orange circles).
For $E_\text{pot}^{(2)}$, this difference can be easily explained by the different magnitudes of the lattice charge and spin susceptibilities which have been analyzed in Sec.~\ref{sec:susceptibilities}.
In fact, $\cdensF$ is smaller for \lamdm than for \lamm while the opposite behavior is observed for $\cmagnF$ which is larger for \lamdm compared to \lamm.
Equation~(\ref{equ:lambdacondition2}) for $E_\text{pot}^{(2)}$ then implies that the corresponding potential energy for \lamdm is smaller compared to the one obtained with \lamm.

The fact, that the potential energy of \lamdm is smaller than $E_\text{pot}^{(1)}$ of \lamm can be easily understood by the difference in the size of the self-energies in both approaches.
Considering Eq.~(\ref{equ:lambdacondition2}) for the calculation of $E_\text{pot}^{(1)}$, we find that
\begin{align}
\label{equ:epot1rewrite}
        E_\text{pot}^{(1)}
    &=
        \sum_{\nu\bk}\Sigma^\nu_\bk G^\nu_\bk
        =
        \sum_{\nu\bk} \left( \frac{i\nu+\mu-\varepsilon_\bk}{i\nu-\varepsilon_\bk-i\text{Im}\Sigma^{\nu}_\bk}-1 \right) \nonumber\\
    &=
        \frac{U}{4}
        + \sum_{\nu\bk}  \left(
        \frac{
            i\nu-\varepsilon_\bk
        }{
            i[\nu-\text{Im}\Sigma^{\nu}_\bk]-\varepsilon_\bk
        }-1 \right).
\end{align}
In the second line, we have used that $\mu=\frac{U}{2}=\text{Re}\Sigma^\nu_\bk$ for the particle-hole symmetric case at half filling ($n=1$) and the dispersion relation $\varepsilon_\mathbf{k}$ is defined as the Fourier transform of the hopping matrix. 
Considering that $\text{Im}\Sigma_\bk^\nu<0$ for $\nu>0$~\footnote{Note that violations of the analyticity condition $\text{Im}\Sigma_\bk^\nu<0$ for $\nu>0$ have been found in ladder D$\Gamma$A calculations for the one-dimensional Hubbard model for momenta far away from the Fermi level in Ref.~\cite{Valli2015}. We do not observe such violations in three dimensions for both version of the ladder D$\Gamma$A.}, it is obvious that a larger self-energy will lead to a smaller potential energy $E_\text{pot}^{(1)}$ (and vice versa). 
% (at least for the most relevant momenta $\mathbf{k}_\text{F}$ at the Fermi level as well as in the asymptotic high-frequency regime)
As we have discussed in Sec.~\ref{sec:resultssigma}, the self-energies for \lamdm are indeed larger than the corresponding self-energies for \lamm in the entire parameter regime which explains the corresponding differences in the potential energies.
\begin{figure*}[t!]
    \centering
    \includegraphics[width=0.46\linewidth]{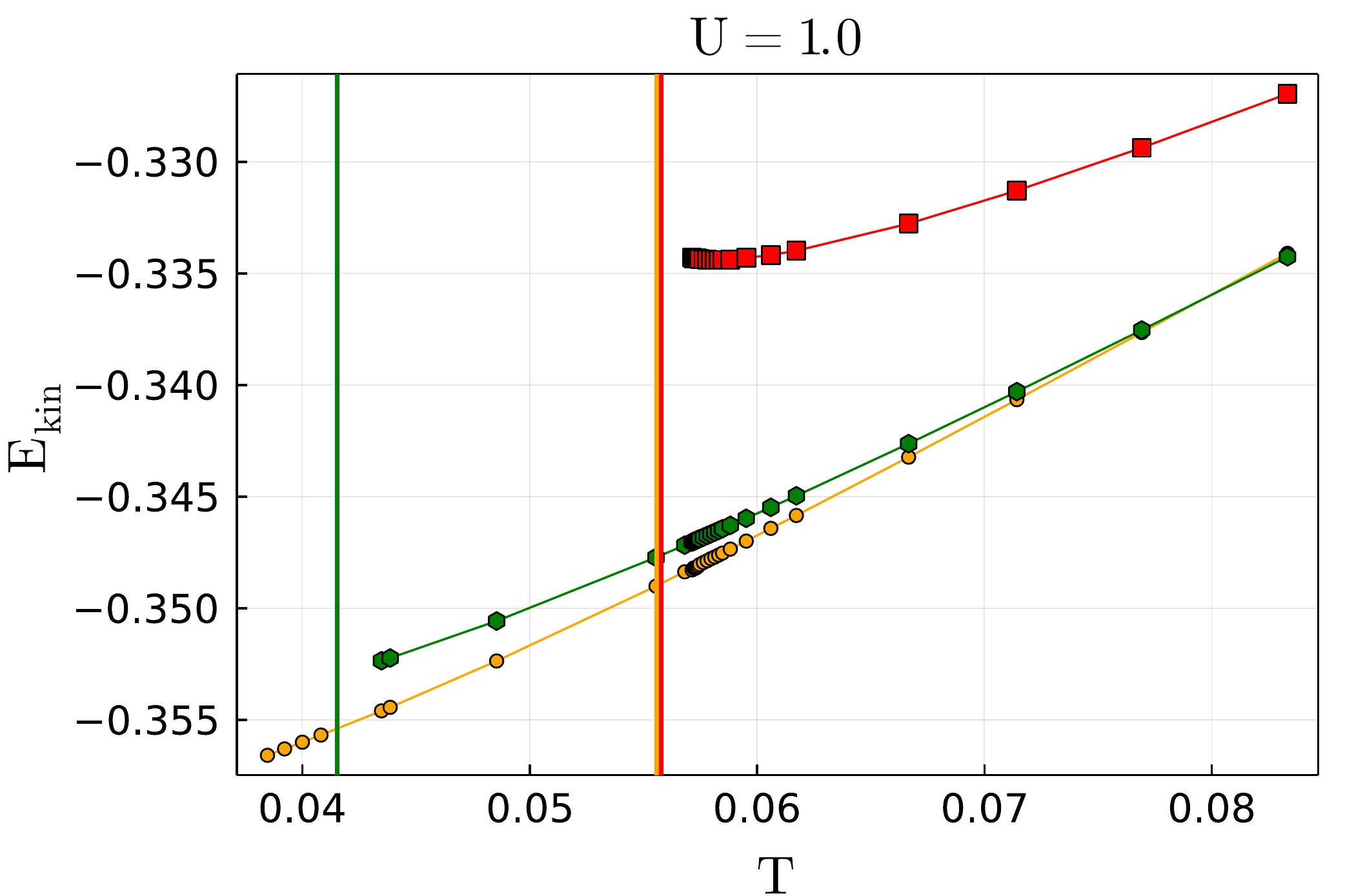}
    \includegraphics[width=0.46\linewidth]{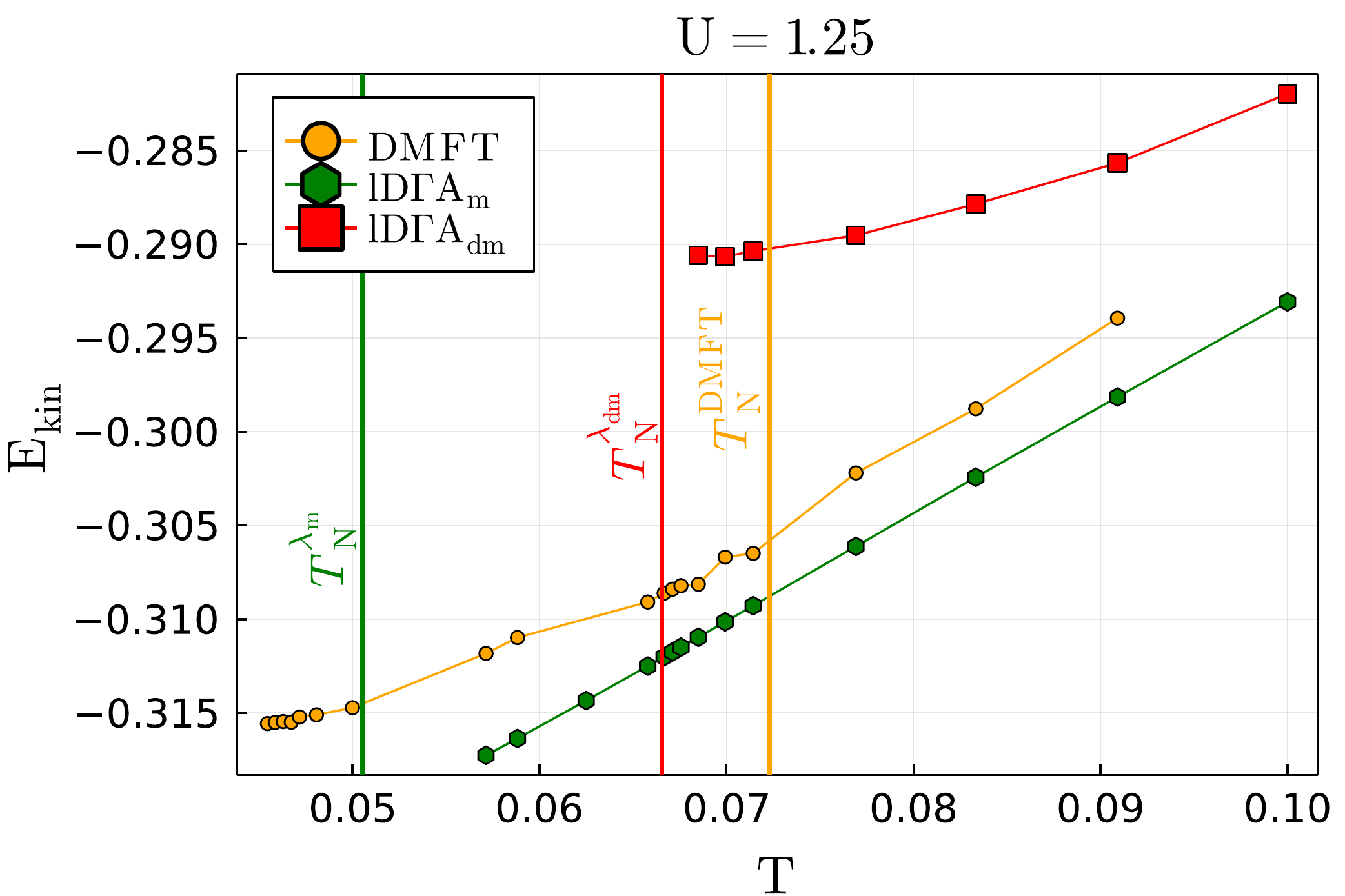}\hfill
    \includegraphics[width=0.46\linewidth]{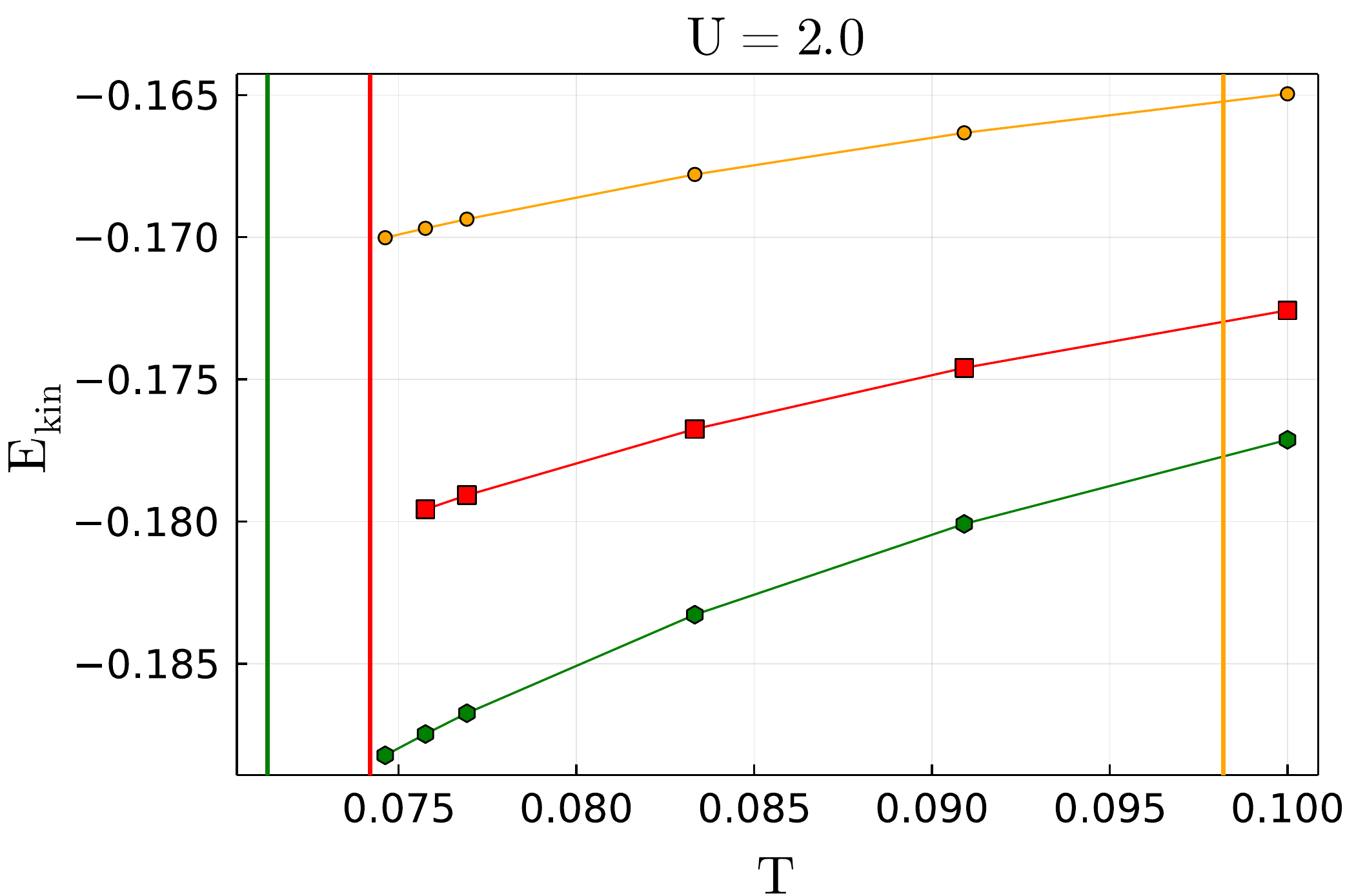}
    \includegraphics[width=0.46\linewidth]{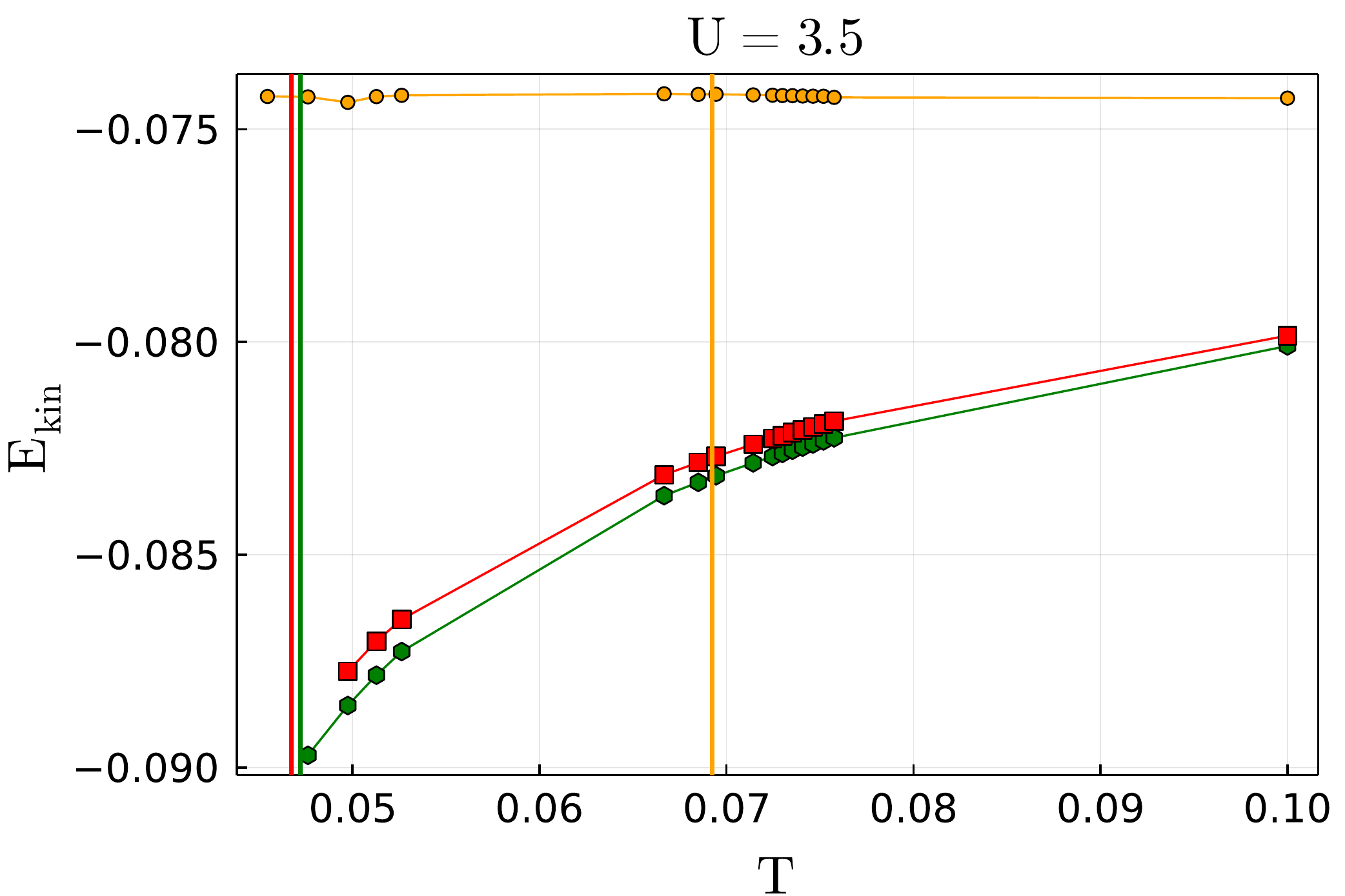}
    \caption{Kinetic energy as a function of temperature for four different values of $U$ as obtained by the DMFT (orange circles), \lamm (green hexagons), and \lamdm (red squares). Vertical lines indicate the transition temperature $\TN$ of the respective method.}
    \label{fig:ekin}
\end{figure*}

Let us also briefly comment on the DMFT potential energy $E_\text{pot}^{(2)}$ (brown triangles in Fig.~\ref{fig:epot}) which is obtained from the charge and spin susceptibilities $\cdensF$ and $\cmagnF$ without any $\lambda$-correction via the left-hand side of Eq.~(\ref{equ:lambdacondition2}). 
We observe that $E_\text{pot}^{(2)}$ of DMFT is smaller than the corresponding $E_\text{pot}^{(2)}$ for the \lamm approach. 
This can be understood by the fact that $\cmagnF$ is larger for DMFT compared to \lamm (c.f., orange circles and green hexagons in the upper panels in Fig.~\ref{fig:suscomega}) and enters into Eq.~(\ref{equ:lambdacondition2}) for $E_\text{pot}^{(2)}$ with a negative sign.

The situation is more complicated for \lamdm where an additional renormalization via $\ldens>0$ is applied to $\cdensF$. 
Such a correction has the opposite effect compared to the $\lambda$-correction in the spin channel because the charge correlation function enters in $E_\text{pot}^{(2)}$ with a positive sign. 
Hence, the introduction of $\ldens$ tends to suppress the corresponding potential energy with respect to DMFT while $\lmagn$ typically increases it.
At weak coupling ($U=1$ and $U=1.25$, upper panels in Fig.~\ref{fig:epot}), we can see that the renormalization of the charge susceptibility has indeed a more pronounced effect and the $E_\text{pot}^{(2)}$ of DMFT (brown triangles) are larger than the corresponding results for \lamdm (red squares).
On the contrary, at strong coupling ($U=2$ and $U=3.5$, lower panels in Fig.~\ref{fig:epot}) the reduction of the potential energy due to the spin susceptibility dominates and $E_\text{pot}^{(2)}$ of DMFT is located below the corresponding \lamdm result (Note that for $U=2$ only one temperature point is located above $\TN$ of DMFT where $E_\text{pot}^{(2)}$ is well defined in DMFT. The corresponding value is not shown in the figure but explicitly given in the caption.).
In fact, at these large values of $U$, $E_\text{pot}^{(2)}$ of DMFT becomes even negative due to the large value of the unrenormalized spin susceptibility.
This unphysical behavior at strong coupling has already been reported for the two-dimensional Hubbard model in Ref.~\onlinecite{vanLoon2016}. 
In the heat map for $E_\text{pot}^{(2)}$ of DMFT in Fig.~\ref{fig:Epot_DMFT_2_grid}, it can be clearly seen that such unphysical negative values emerge at $U\approx2$. 
Note also, that below $\TN$ of DMFT $E_\text{pot}^{(2)}$ becomes numerically ill defined, since the divergence of the magnetic susceptibility around the antiferromagnetic ordering vector introduces $\bk$-sampling dependent noise.

Let us finally comment on the temperature dependence of $E_\text{pot}$ for the different methods.
At weak coupling ($U=1$ and $U=1.25$, upper panels in Fig.~\ref{fig:epot}), we observe an increase of $E_\text{pot}^{(1)}$ for DMFT (orange circles) and $E_\text{pot}^{(1)}$ as well as $E_\text{pot}^{(2)}$ for \lamm (green hexagons and blue diamonds) upon decreasing temperature.
For DMFT this is indeed the expected behavior as the system becomes more metallic at low temperatures.\footnote{Note that the one-particle as well as the impurity quantities of DMFT are not affected by nonlocal antiferromagnetic fluctuations in the proximity of the antiferromagnetic phase transition} 
For the \lamm results, on the other hand, this increase with decreasing temperature is unphysical as the double occupancy is expected to become smaller when approaching the antiferromagnetic order at $\TN$.
The latter (physical) behavior is indeed observed when we consider $\ldens>0$ in the \lamdm approach (red squares) which further demonstrates the improved consistency of our new approach with respect to the \lamm method. At intermediate and strong coupling ($U=2$ and $U=3.5$, lower panels) both versions of D$\Gamma$A feature the physically correct decrease of $E_\text{pot}^{(1)}$ with decreasing temperature. 

Let us close this section by briefly discussing the kinetic energy of the system as depicted in Fig.~\ref{fig:ekin}. It is calculated via the equation
\begin{equation}
\label{equ:ekin}
        E_\text{kin}
    =
        \sum_{\nu\mathbf{k}}\varepsilon_\bk G^\nu_\bk.
\end{equation}
At weak coupling, the antiferromagnetic state is of Slater type. 
As it has been discussed extensively in Ref.~\onlinecite{Rohringer2016}, this implies that the symmetry broken phase is stabilized by a reduction of the potential energy while the kinetic energy is larger in the symmetry broken than in the normal state.
It has been demonstrated in the latter publication, that at the very small value of $U=0.75$, this also holds for the corresponding fluctuations above $\TN$ and is reflected in the kinetic energy of \lamm with respect to the corresponding DMFT result.
We also find this behavior for both the \lamm and \lamdm at $U=1$ in the left upper panel of Fig.~\ref{fig:ekin}. 

At a slightly larger value of the coupling ($U=1.25$, right upper panel of Fig.~\ref{fig:ekin}),  $E_\text{kin}$ for \lamm (green hexagons) is located below the corresponding DMFT result (orange circles).
This implies that within the \lamm method the system is already in an intermediate coupling region.
On the contrary, our new result for \lamdm (red squares) predicts a kinetic energy above the one of DMFT indicating that the system is still in the weak coupling regime. 
This is indeed consistent with the corresponding $\TN$ which is very close to the DMFT result at this value of $U$. Overall one can see that the \lamdm approach extends the range where a Slater type antiferromagnetism is observed with respect to the \lamm method.

At intermediate and strong coupling, both D$\Gamma$A approaches predict a kinetic energy below the one of DMFT which is the expected behavior in this parameter regime. 
Interestingly, $E_\text{kin}$ for \lamdm is always larger (smaller in absolute value) than the one obtained via \lamm. 
This is again a consequence of the larger self-energy in the \lamdm approach and can be explicitly demonstrated by rewriting Eq.~(\ref{equ:ekin}) for $E_\text{kin}$ into a similar form as the equation for $E_\text{pot}^{(1)}$ in Eq.~(\ref{equ:epot1rewrite}).
The difference between $E_\text{kin}$ of the two versions of ladder D$\Gamma$A decreases upon increasing $U$ as the charge renormalization becomes gradually less important and eventually almost vanishes at the strongest coupling $U=3.5$ (see also the corresponding discussions in the previous sections).

Let us finally mention, that the kinetic energy is also accessible from the two-particle charge and spin susceptibilities via the $f$-sum rule.
It states that the coefficients of the $\frac{1}{(i\omega)^2}$ tails of these correlation functions are related to this thermodynamic observable (see, e.g., Refs.~\cite{Krien2017,Rohringer2016}). 
For DMFT this leads to the same results for $E_\text{kin}$ as the calculation at the one-particle level via Eq.~\ref{equ:ekin}, because DMFT is a conserving theory. 
However, since the $\lambda$-correction in its present form does not change the asymptotic $\frac{1}{(i\omega)^2}$ contribution of $\chi_{\text{d},\mathbf{q}}^{\lambda_\text{d},\omega}$ and $\chi_{\text{m},\mathbf{q}}^{\lambda_\text{m},\omega}$, the D$\Gamma$A results for $E_\text{kin}$ at the two-particle level coincide with the corresponding DMFT data and are, hence, not equivalent to the D$\Gamma$A results obtained from Eq.~(\ref{equ:ekin}).

\section{Conclusions and Outlook}\label{sec:conclusions}

In this paper, we have presented a method which takes into account nonlocal correlations beyond the local ones of DMFT and fulfills specific exact sum rules which connect one- and two-particle correlation functions.
Our new approach is based on the ladder dynamical vertex approximation where nonlocal corrections to the purely local DMFT self-energy are obtained via a diagrammatic expansion around DMFT.
More specifically, within this method a momentum dependent self-energy is constructed from the DMFT Green's function and the DMFT charge and spin lattice susceptibilities. 
Since we do not perform a fully self-consistent treatment of the problem the results initially violate certain sum rules for these susceptibilities which control the total density and the potential energy of the system. 
To overcome this problem, we have introduced a mass renormalization of the charge and spin susceptibilities by means of (constant) parameters $\ldens$ and $\lmagn$ which are determined by the requirement that the above mentioned sum rules are fulfilled. 
A simpler version of this idea, where a correction is applied only to the spin channel, has been already successfully exploited in previous research works~\cite{Katanin2009,Rohringer2016}.

We have applied our new approach to the three-dimensional half-filled Hubbard model on a simple cubic lattice with nearest neighbor hopping which features an antiferromagnetically ordered phase at low temperatures for all values of the interaction strength. 
The introduction of the correction parameters $\ldens$ and $\lmagn$ leads to a mutual renormalization of charge and spin fluctuations which can be usually only achieved in far more complicated theories such as the parquet approach.
The latter is, however, restricted to simple one-band models due to its numerical complexity while our method is, in principal, applicable for multi-band systems or systems with a nonlocal interaction.

We have demonstrated that our method, which takes into account the renormalization of both the charge and spin susceptibility, improves several results compared to the above mentioned previous version of D$\Gamma$A where only the spin susceptibility has been renormalized. 
In particular, at weak-to-intermediate coupling it predicts a higher transition temperature $\TN$ to the antiferromagnetically ordered state with respect to the old approach which is in good agreement with dual fermion and diagrammatic Monte Carlo calculations. 
At strong coupling it gradually becomes similar to the old technique as charge fluctuations are strongly suppressed and their renormalization has (almost) no effect on the physical results.

We have also analyzed the potential energy which is obtained by our new method. 
In contrast to DMFT and the previous version of D$\Gamma$A, it is uniquely defined and lower than the corresponding ones obtained by the latter approaches. 
Moreover, at weak coupling it always decreases upon decreasing temperatures approaching the antiferromagnetic phase transition which is indeed the expected behavior for a Slater type antiferromagnet where the ordered phase is stabilized by the potential energy. 
The kinetic energy is located above the one of DMFT at weak coupling which is also consistent with Slater type antiferromagnetism while the previous non-self-consistent version of D$\Gamma$A predicts a lower kinetic energy with respect to DMFT. 
Overall, our new approach describes the weak and intermediate coupling regime, where charge fluctuations still play an important role, substantially better than the previous approach where a renormalization of charge fluctuations is absent.

Let us finally state, that our new method is not yet fully two-particle consistent as it violates sum rules which originate from conservation laws such as the $f$-sum rule.
The inclusion of such consistency relations in the D$\Gamma$A formalism (and also other diagrammatic extension of DMFT) is an interesting future research perspective (for some preliminary ideas see, e.g., Ref.~\cite{Rohringer2016}).
%Furthermore, we have not presented results out of half filling in order to draw a clearer picture of the differences between D$\Gamma$A$_\mathrm{m}$ and D$\Gamma$A$_\mathrm{dm}$. 

In this work, we have presented data only at half-filling to avoid additional technical expenses for the determination of the chemical potential for a given average particle density.
However, there are no restrictions on the method preventing a generalization to finite doping. In fact, we expect that the  numerical accuracy even improves in this situation due to an increase of charge fluctuations.
Finally, the extension of the presented approach to more realistic multi-orbital systems can potentially lead to an improved theoretical description of nonlocal correlation effects in real materials.

\paragraph*{Acknowledgements}
We thank K. Held, A. Katanin, Th. Sch\"afer and A. Toschi for useful discussions. We acknowledge financial support from the Deutsche Forschungsgemeinschaft  (DFG)  through Projects No.~407372336 and No.~449872909. The work was supported by the North-German Supercomputing Alliance (HLRN).

\appendix

\section{Use of symmetries}\label{app:symm}

\begin{figure}[t!]
    \centering
    \includegraphics[width=\linewidth]{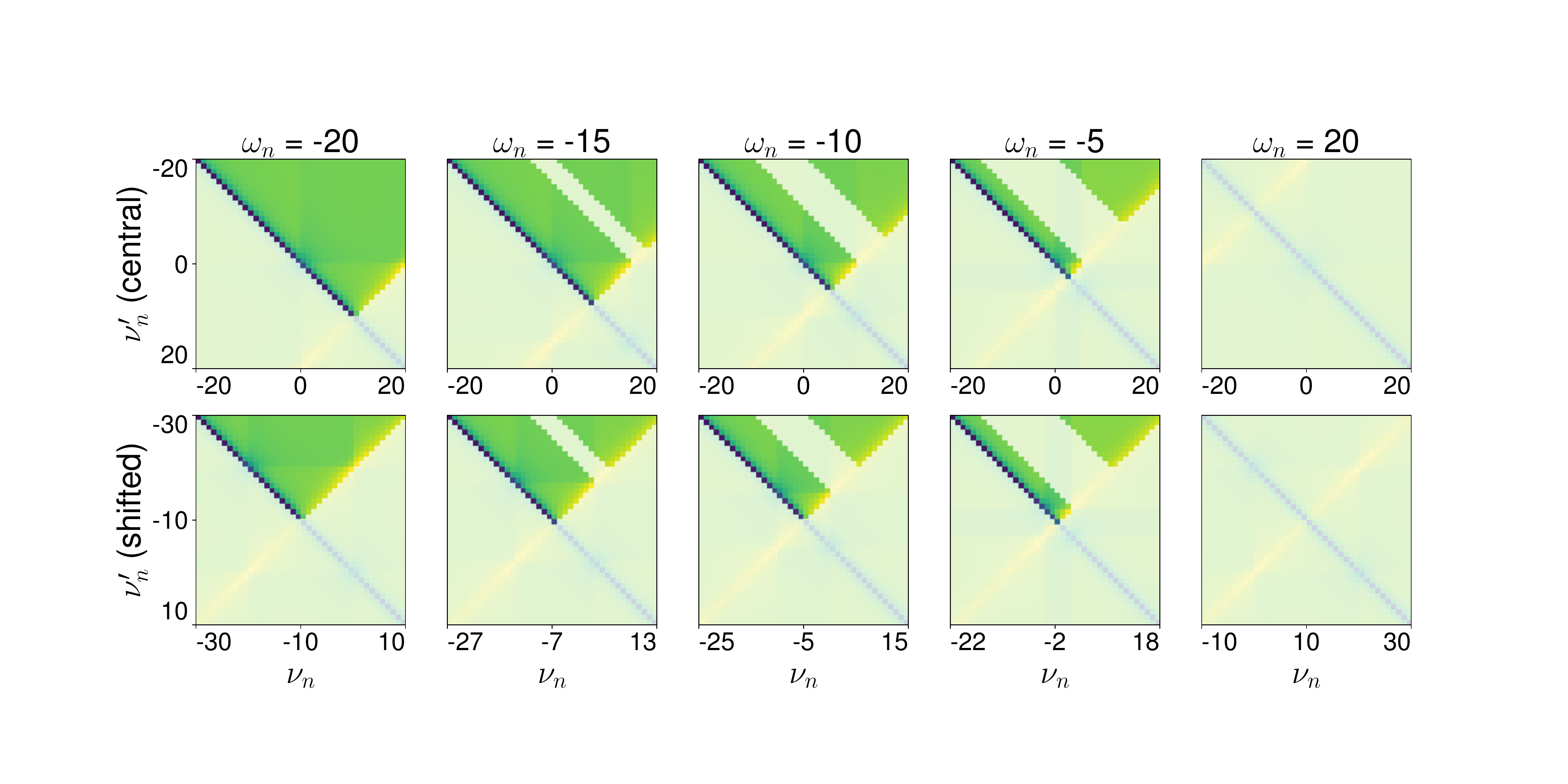}
    \caption{Full vertex $F^{\nu\nu'\omega}_\mathrm{m}$ as a function of $\nu, \nu'$ at 5 different $\omega$. Labels indicate the (integer) index of the Matsubara frequency.
    Shaded colors indicate points connected to others by symmetries. (Top) Unshifted grids. (Bottom) $\nu,\nu'$ shifted by $-\omega/2$}.
    \label{fig:grid_red}
\end{figure}

The irreducible vertex $\Gamma_r^{\nu\nu'\omega}$, which is required for the calculation of the lattice generalized susceptibility $\chi_{r,\mathbf{q}}^{\nu\nu'\omega}$ in Eq.~(\ref{equ:chi_gen_def}), is calculated from the local generalized susceptibility $\chi_r^{\nu\nu'\omega}$ via a local version of Eq.~(\ref{equ:chi_gen_def}). 
The latter is obtained directly from the ED impurity solver which represents the numerically most expensive part of the entire method. To reduce the number of frequencies for which $\chi_r^{\nu\nu'\omega}$ has to be evaluated explicitly we have considered two algorithmic improvements: 

(i) We have shifted the fermionic Matsubara frequencies $(\nu,\nu')$ by $-\frac{\omega}{2}$.
This improves our calculations since the main nonperturbative structures of $\chi_r^{\nu\nu'\omega}$ are centered around $(-\frac{\omega}{2},-\frac{\omega}{2})$ (see Refs.~\onlinecite{Rohringer2012,Rohringer2013a}). 
This can be seen in the upper panels of Fig.~\ref{fig:grid_red} where we present the full vertex $F^{\nu\nu'\omega}_\mathrm{m}$, which is obtained from $\chi_\text{m}^{\nu\nu'\omega}$ by amputating the four outer Green function lines~\cite{Rohringer2012}, as a function of $\nu,\nu'$ at different $\omega$ slices. The main structures, indicated by the crossing of the yellow and blue diagonal contributions, indeed move to the center if the frequency grid is shifted by $-\frac{\omega}{2}$.

(ii) We have considered all physical symmetries of $\chi_r^{\nu\nu'\omega}$ which allows us to reduce the actual calculation to a subset of Matsubara frequencies in the selected frequency grid. In fact, the generalized susceptibility $\chi_r^{\nu\nu'\omega}$ is equivalent for all frequency triples $(\nu,\nu',\omega)$ which are related to each other via a specific physical symmetry. 
Consequently, it is sufficient to determine this correlation function for only {\em one} of this related triples. 
In Fig.~\ref{fig:grid_red}, the frequencies which are related to another one by a symmetry are shaded.
Overall this leads to a reduction in the number of triples $(\nu,\nu',\omega)$ by more than a factor of~10.

For the determination of equivalent arguments for the two-particle Green's function, we have proceeded in the following way:
\begin{enumerate}
    \item Define the grid of size $N_\mathrm{tot} = N^2_\nu \times N_\omega$, possibly with shifted $\nu,\nu'$ values.
    \item Define the symmetry mapping, i.e.,~$f(p) = $ ``\textit{list of points $p$ maps to.}''. This includes only direct symmetries, so for 5 symmetries in the system, the list will have the length 5.
    \item Construct an undirected graph with $N_\mathrm{tot}$ vertices, each representing a point on the grid while edges for each vertex $v$ are given by $f(v)$. It suffices to loop over all vertices and call $f$ on its value, disregarding double edges. In case the equivalent points are related by some operation other than the identity, for example complex conjugation, one has to track the operation connecting two vertices as edge ``weight,'' for example, in a parent array.
    \item Determine all connected components, for example using depth first search~\cite{Even2012}, and choose a (random) node as representative. The construction of the mapping from the reduced to the full grid, including the ``weights,'' can be done with a modified depth first search as well.
    \item Hand off the reduced grid and mapping to the full grid to the impurity solver and D$\Gamma$A code.
\end{enumerate}

\section{Improved asymptotics}
\label{app:numericaldetails}

\begin{figure}[t!]
    \centering
    \includegraphics[width=\linewidth]{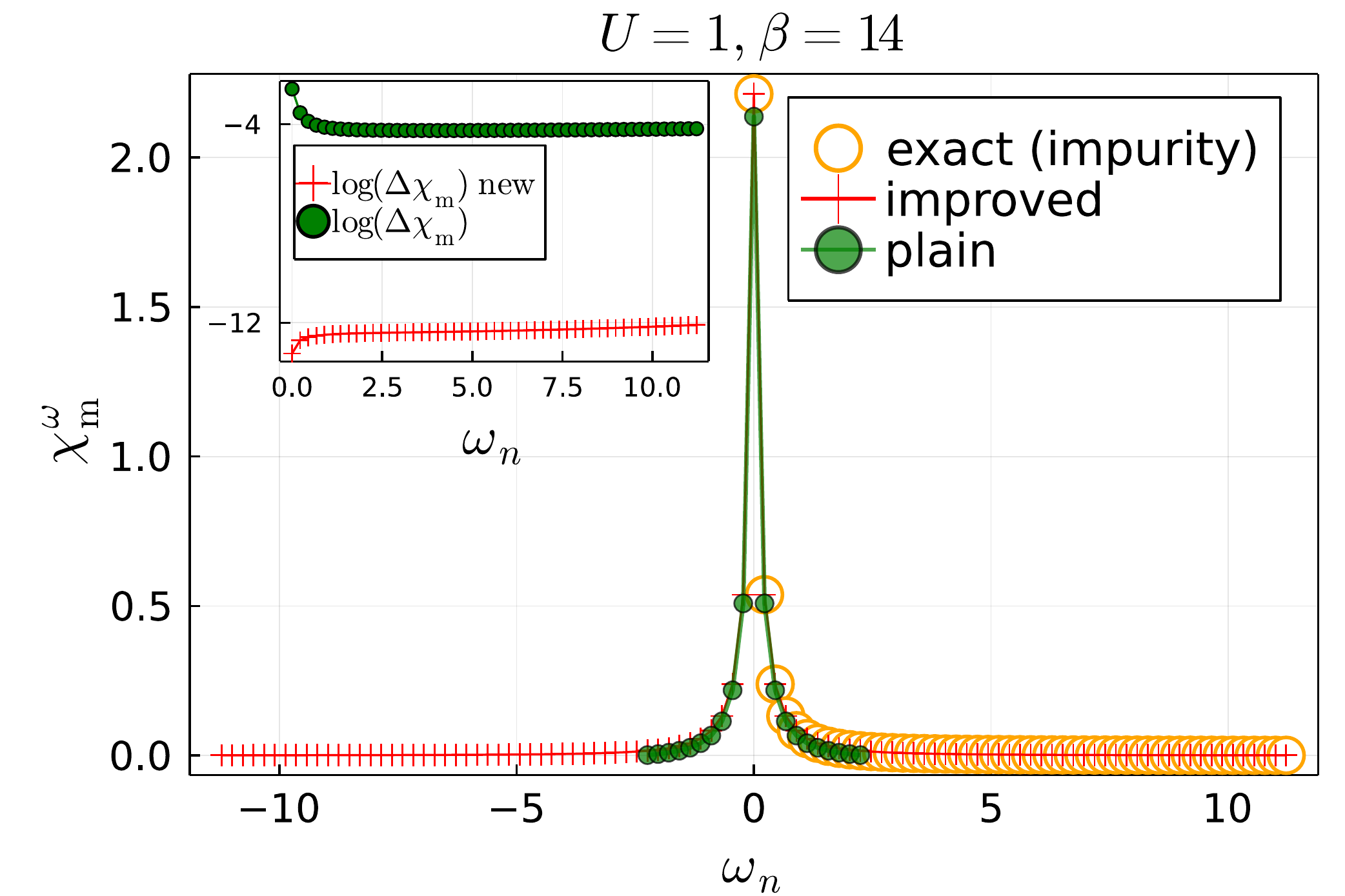}
    \caption{Local physical spin susceptibility $\chi_\text{m}^\omega$ as a function of $\omega$ for $U=1$ and $\beta=14$ as obtained by summing the generalized physical susceptibility $\chi_\text{m}^{\nu\nu'\omega}$ over the fermionic frequencies $\nu$ and $\nu'$ using the improved frequency sums (red crosses) versus a plain frequency sum on a finite grid (green filled circles) and the exact results obtained directly from the impurity solver (empty orange circles). The inset shows the logarithm of the difference between normal summation and improved summation to the exact values. In the main panel, the plain Matsubara summation is cut of at values of $\omega$ where it becomes negative (i.e.,~unphysical).}
    \label{fig:chi_after_bse_impr}
\end{figure}
For the \lamdm method, it is necessary to solve the two coupled equations in Eq.~(\ref{equ:lambdacondition1}) and Eq.~(\ref{equ:lambdacondition2}) for $\lmagn$ and $\ldens$ simultaneously. 
This requires a precise evaluation of the corresponding frequency sums in these relations as well as in Eqs.~(\ref{equ:chi_phys_def}) and (\ref{equ:gamma_def}). For a numerical evaluation, these sums over an infinite number of Matsubara frequencies obviously have to be restricted to a finite frequency grid. 
A plane summation in such a finite frequency domain is typically not sufficient to determine $\ldens$ and $\lmagn$, but the consideration of the high-frequency asymptotic behavior of all involved correlation functions provides accurate and stable enough results, even if small frequency grids or noisy Monte Carlo input data is used.
This section gives an overview of the improved Matsubara summation method used to perform the Matsubara frequency sums for the determination of the $\lambda$ parameters.

The central idea for this summation technique is to divide functions of Matsubara frequencies $f^\omega$, where $\omega$ is now a generic variable which can represent a (set of) fermionic and/or bosonic Matsubara frequencies, into a core and asymptotic region.
The summation is then performed separately for both.
In the core region (indicated by a subscript ``c''), a sum over the exact numerical data for $f^\omega$ is carried out while in the asymptotic high-frequency regime $f^\omega$ (or shell region, indicated by a subscript ``s'') is replaced by leading order diagrams $T^\omega$ which do not decay as a function of frequency. The frequency sum over the latter can be performed semi-analytically. Formally, this idea can be represented as follows:
\begin{align*}
    \sum_\omega f^\omega
      &= \SumCoreOmega \left( f^\omega - T^\omega + T^\omega \right) + \SumAsymOmega \left( f^\omega - T^\omega + T^\omega \right)\\
      &\approx \SumCoreOmega \left( f^\omega - T^\omega \right) + \sum_\omega T^\omega,
\end{align*}
where in the second line of this equation we have neglected the term $f^\omega-T^\omega$ in the high-frequency shell region as it rapidly decays with increasing frequency $\omega$. As discussed in Appendix~\ref{app:symm}, the core region $\Omega_\text{c}=\Omega^\nu_\text{c}$ for the bosonic or fermionic frequency $\omega$ can depend on a (fermionic or bosonic) Matsubara frequency $\nu$ which is not involved in the summation when shifted grids are considered for the calculation.
If two indices are used for a core or shell summation, the notation implies that the \textit{tuple} lies either within the core or  the shell region.

In the following we will denote quantities summed over the core and shell region by corresponding superscripts. 
Furthermore, we use a tilde to distinguish quantities obtained through improved summation from ones obtained through plain summation.
For the bare susceptibility (``bubble'' term), this reads as follows:
\begin{align*}
    \chi^{\nu\omega}_{0,\bq} 
    &=
    -
    \beta\sum_\bk  G^{\nu}_\bk G^{\nu + \omega}_{\bk + \bq}  \nonumber\\
    \chi^\omega_{0,\bq},
    &=
    \SumCoreNu \chi^{\nu\omega}_{0,\bq} 
        +
        \SumAsymNu \chi^{\nu\omega}_{0,\bq} 
     = \chi^{\mathrm{core},\omega}_{0, \bq} + \chi^{\mathrm{shell},\omega}_{0, \bq}.\label{equ:defchi0nu}
\end{align*}

The asymptotic contribution to the bare susceptibility $\chi^{\omega,\mathrm{shell}}_{0, \bq}$ can be obtained directly from the high-frequency tails of the DMFT Green's functions.
Following the derivation of the high-frequency behavior of the vertex functions $F_{r,\mathbf{q}}^{\nu\nu'\omega}$ and $\Gamma_r^{\nu\nu'\omega}$ as discussed in Refs.~\onlinecite{Rohringer2012,Tagliavini2018,Wentzell2020}, we obtain for the improved frequency sums for the calculation of $\chi_{r,\mathbf{q}}^\omega$ and $\gamma_{r,\mathbf{q}}^{\nu\omega}$ in Eqs.~(\ref{equ:chi_phys_def}) and (\ref{equ:gamma_def}) the following expressions:
\begin{align}
    \tilde{\lambda}^{\nu\omega}_{r,\bq} 
  & =  
    \left( 1 \mp U \chiZeroShell \right)^{-1} \cdot 
    \left( 
        \lambdaFullCore
        + \lambdaFullShell 
    \right) \\
    \tilde{\chi}^{\omega }_{r,\bq} 
  & = 
    \left( 
        1 - \left( U
            \chiZeroShell  
        \right)^2 
    \right)^{-1} \cdot
    \left(  
        \chiFullCore
        + \chiFullShell
    \right)\\
  \tilde{\gamma}^{\nu\omega}_{r,\bq}
  & =
  \frac{1 \mp \tilde{\lambda}^{\nu\omega}_{r,\bq} }{1 \pm U \tilde{\chi}^{\omega }_{r,\bq}}.
\end{align}
We have introduced the following abbreviations:
\begin{align*}
   F^{\nu\nu'\omega }_{\text{m},\text{diag}} 
  & = 
  \frac{1}{2} \chi^{\nu - \nu'}_\text{d} 
  - \frac{1}{2} \chi^{\nu-\nu'}_\text{m} 
  + \chi^{\nu+\nu'+\omega}_{\text{pp},\uparrow\downarrow}  \\
    F^{\nu\nu'\omega }_{\text{d},\text{diag}} 
        & = \frac{1}{2} \chi^{\nu-\nu'}_\text{d} + \frac{3}{2} \chi^{\nu-\nu'}_\text{m} - \chi^{\nu+\nu'+\omega}_{\text{pp},\uparrow\downarrow}\\
   \lambdaFullCore
  & =
    \pm \beta \SumCoreNup \left( 
        \delta_{\nu \nu'}
        - \frac{\chi^{\nu\nu'\omega}_\bq}{\chi^{\nu\omega}_{0,\bq}}
    \right)
    \\
   \lambdaFullShell
  & =  
    U \chiZeroShell
    \mp U^2 \SumAsymNup 
        F^{\nu\nu'\omega }_{r,\text{diag}} \chi^{\nu'\omega}_{0,\bq}
    \\
   \chiFullCore 
  & =  
    \SumCoreNuNup 
        \chi^{\nu\nu'\omega}_{r,\bq}    \\
    \chiFullShell
  & = \pm U \left(\chiZeroShell\right)^2 
%   & \qquad
  - U^2 \SumAsymNuNup
        \chi^{\nu\omega}_{0,\bq} 
        F^{\nu\nu'\omega }_{r,\text{diag}} 
        \chi^{\nu'\omega}_{0,\bq} \\
  & \qquad
  - \SumAsymNup \chi^{\nu'\omega}_{0,\bq}
  \left({
    1 
    + 2U \SumCoreNu
        (\tilde{\lambda}^{\nu\omega}_{r,\bq} \mp 1) \chi^{\nu\omega}_{0,\bq}
  }\right).
\end{align*}
The local physical susceptibilities $\chi_r^\omega$ for the diagonal term $F^{\nu\nu'\omega }_{r,\text{diag}}$ of the full vertex are local and can be obtained directly from the DMFT impurity solver.
The contributions from $F^{\nu\nu'\omega }_{r,\text{diag}}$ are typically negligible, when the physical susceptibilities in the three channels fall off sufficiently fast with increasing frequency.
%Defining $F^{\nu\nu'\omega }_{r,\text{diag}} \equiv 0$ substantially simplifies the expressions above and can therefore be considered in order to increase numerical performance.
Fig.~\ref{fig:chi_after_bse_impr} shows a benchmark for our improved summation method for the summation of the purely local generalized susceptibility $\chi_\text{m}^{\nu\nu'\omega}$ (in the spin channel) over the fermionic Matsubara frequencies $\nu$ and $\nu'$ yielding the physical susceptibility $\chi_\text{m}^\omega$. 
The $\chi_\text{m}^\omega$ obtained via our improved summation technique (red crosses) agrees excellently with the exact results (orange empty circles) obtained directly from the impurity solver. 
The $\chi_\text{m}^\omega$ obtained through a plain sum on a finite frequency grid (green filled circles), on the other hand, shows substantially larger deviations from the exact results (see inset in Fig.~\ref{fig:chi_after_bse_impr}) and eventually features unphysical negative values at larger frequencies $\omega$ which is not observed for the improved frequency sum.

\section{Root finding procedure}\label{sec:root_finding}

\begin{figure}[t!]
    \centering
    \vspace*{1.5mm}
    \includegraphics[width=\linewidth]{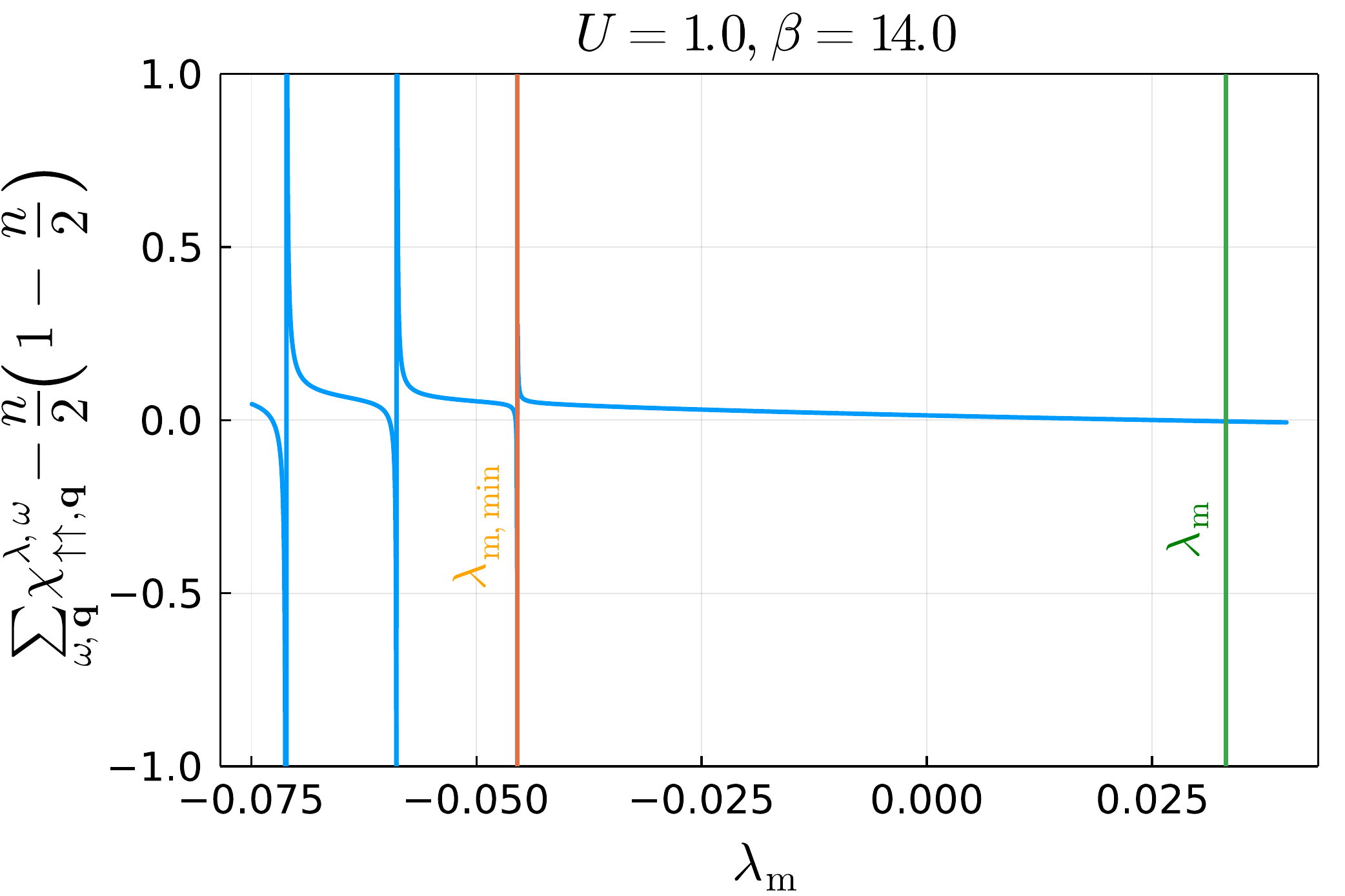}
    \caption{$\sum_{\omega\bq} \chi^{\lambda,\omega}_{\mathrm{\uparrow\uparrow},\bq} - \frac{n}{2}(1 - \frac{n}{2}) = \frac{1}{2} \sum_{\omega\bq}\left(\chi^{\ldens,\omega}_{\mathrm{d},\bq} + \chi^{\lmagn, \omega}_{\mathrm{m},\bq}\right) - \frac{n}{2}(1 - \frac{n}{2})$ for $\ldens = 0$ showing the root for the equation (which determines $\lmagn$) and poles for $\lmagn < \lambda_\mathrm{m,min}$.}
    \label{fig:chi_lambda_spikes}
\end{figure}

Finding the roots $\ldens$ and $\lmagn$ for Eqs.~(\ref{equ:lambdacondition1}) and~(\ref{equ:lambdacondition2}) can be done with any root finding algorithm, but requires two considerations in order to yield reasonable results:
(i) the physical tails must be free of finite size effects and
(ii) the unphysical, $\bq$ dependent poles of Eq.~(\ref{equ:lambdacorr}) must be avoided in the calculation (i.e.~the resulting $\lambda$ corrected physical susceptibility must non-negative).

The first difficulty can be overcome by using the procedure for treating the high-frequency asymptotic tails of all correlation functions as explained in Appendix~\ref{app:numericaldetails}.
The second requirement has been avoided by the procedure described in Sec.~\ref{sec:determinationlambda}.
However, this method is not applicable for large simulations, due to the numerical cost of the many evaluations of the equation of motion, required to obtain the curves in Figs.~\ref{fig:lambda_sp_of_ch_res} and \ref{fig:c2_curved_3U} over a large range of $\lambda$ values.
Instead, a more elaborate root finding algorithm, such as the Newton method, is preferable requiring substantially fewer evaluations of the EOM. 
To avoid unphysical solutions, let us consider that the $\lambda$ corrected physical susceptibilities are continuous and monotonically decreasing (as a function of $\lambda$) for all $\lambda$ values larger than the largest pole, i.e.,~$\lambda_r > \lambda_{r,\text{min}}$~\cite{Schaefer2016}.
The location of the largest pole is:
\begin{align*}
    \lambda_{r,\mathrm{min}} & = \min_\bq \frac{1}{\chi^{\omega=0}_{r,\bq}}
\intertext{The monotonously decreasing behavior follows directly from the derivative of $\chi_{r,\mathbf{q}}^{\lambda_r,\omega}$ with respect to $\lambda_r$:}
   \frac{\diff }{\diff \lambda_r} \sum_{\omega\bq} \chi^{\lambda_r,\omega}_{r,\bq} & = -\sum_{\omega\bq} \left( \chi^{\lambda_r,\omega}_{r,\bq} \right)^2 \leq 0
\end{align*}
In Fig.~\ref{fig:chi_lambda_spikes}, the difference between right- and left-hand side of Eq.~(\ref{equ:lambdacondition1}) is shown for $U=1$ and $\beta=14$ as a function of $\lmagn$.
Here we see the $\bq$-sampling dependent divergences for $\lmagn < \lambda_{\mathrm{m},\mathrm{min}}$.
Since both $\lambda$ corrected physical susceptibilities exhibit the same behavior, the $\lambda_{r,\mathrm{min}}$ values can be determined independently.
This pole structure leads to an interval $[\lambda_{r,\mathrm{min}},\infty)$ in which a single root for Eq.~(\ref{equ:lambdacondition1}) is located.
Eq.~(\ref{equ:lambdacondition2}) could in principle exhibit non-monotonous behavior, since the right-hand side is a function of both $\lambda$ corrected physical susceptibilities. 
However, as discussed in Sec.~\ref{sec:determinationlambda} (see Figs.~\ref{fig:lambda_sp_of_ch_res} and \ref{fig:c2_curved_3U}), this does not happen for our calculations. 
Therefore, by means of the following transforming one can obtain a result, guaranteed to yield exactly one, physically correct, root.
\begin{equation}\label{equ:lambda_trafo}
    \tilde{\lambda}_r = \frac{\lambda_{r,\mathrm{min}} - \lambda_{r,\mathrm{max}}}{2} (\tanh{(\lambda_r)} + 1) + \lambda_{r,\mathrm{min}}
\end{equation}
$\lambda_{r,\mathrm{max}}$ can in principle be arbitrarily large, but a reasonable value can be chosen from the known fluctuations strength of the system.
This transformation is then applied to the function, before it is handed over to the solver and the resulting root is transformed in the same way.
%
%    \frac{U}{2}\sum_{\omega\bq}\left(\chi^{\ldens,\omega}_{\mathrm{d},\bq}
%        - \chi^{\lmagn, \omega}_{\mathrm{m},\bq}\right)
%        + U\frac{n^2}{4}\nonumber\\
%    =\underbrace{U \sum_{\omega\bq} \chi^{\bl,\omega}_{\uparrow\downarrow,\bq} 
%        + U\frac{n^2}{4}}_{E^{(2)}_\text{pot}}

For our purposes, we use the multivariate Newton method, which is a reasonable choice due to the low dimensionality of our problem and the smoothness of the search space (see Sec.~\ref{sec:determinationlambda}]).
The Jacobian was determined by finite differences since Eq.~(\ref{equ:lambdacondition2}) involves a convolution over numerical data, making automatic and analytic differentiation challenging.
%This approach is however reasonable in our case, due to the smoothness of the search space, as discussed in Sec.~\ref{sec:determinationlambda}.
Note, that the transformation in Eq.~(\ref{equ:lambda_trafo}) concentrates sampling points at the borders of the search interval.
For very high precision, especially at strong coupling where the Jacobian becomes small (see also Fig.~\ref{fig:lambda_sp_of_ch_res}), one can first run a low precision pass with the transformation and then use the obtained result as a starting point for a high precision search.

%\bibliographystyle{apsrev4-2}
%\bibliography{georg_thesis} 


\end{document}